%% file: neurips_2025.tex
\documentclass{article}

\PassOptionsToPackage{numbers, compress}{natbib}

\usepackage[preprint]{neurips_2025}
\usepackage{makecell}   
\usepackage[utf8]{inputenc} 
\usepackage[T1]{fontenc}    
\usepackage{hyperref}       
\usepackage{url}            
\usepackage{booktabs}       
\usepackage{amsfonts}       
\usepackage{nicefrac}       
\usepackage{microtype}      
\usepackage{xcolor}         
\usepackage{graphicx}       
\usepackage{wrapfig}        
\usepackage{enumitem}       
\usepackage{caption}        
\usepackage[ruled,vlined,linesnumbered]{algorithm2e} 
\usepackage{float}          
\usepackage{titletoc}
\usepackage{listings}       
\usepackage{arydshln}
\usepackage[most]{tcolorbox}
\usepackage{fancyvrb}
\tcbuselibrary{breakable}
\tcbuselibrary{listings}  
\usepackage{pifont}

\newcommand{\xmark}{\ding{55}}   

\usepackage{natbib}

\usepackage{cleveref}
\usepackage{fvextra}
\DefineVerbatimEnvironment{MyVerbatim}{Verbatim}{breaklines=true}

\title{Towards Robust Agentic CUDA Kernel \\Benchmarking, Verification, and Optimization}

\author{%
  Robert Tjarko Lange \\
  Sakana AI\\
  \And
  Qi Sun \\
  Sakana AI \\
  \And
  Aaditya Prasad \\
  Sakana AI \\
  \AND
  Maxence Faldor \\
  Sakana AI \\
  \And
  Yujin Tang \\
  Sakana AI \\
  \And
  David Ha \\
  Sakana AI \\
}

\begin{document}

\maketitle

\begin{abstract}
    Recent advances in large language models (LLMs) demonstrate their effectiveness in scaling test-time compute for software engineering tasks.
    However, these approaches often focus on high-level solutions, with limited attention to optimizing low-level CUDA kernel implementations.
    Additionally, existing kernel generation benchmarks suffer from exploitable loopholes and insufficient diversity in testing conditions, hindering true generalization assessment.
    To address these limitations, we introduce \texttt{robust-kbench}, a new benchmark for rigorous evaluation of kernel performance and correctness across varied scenarios.
    Furthermore, we present a comprehensive agentic framework that automates CUDA kernel discovery, verification, and optimization.
    This pipeline enables frontier LLMs to translate \texttt{torch} code to CUDA kernels and iteratively improve their runtime within our robust evaluation setting.
    Our sequential workflow first translates PyTorch code into equivalent CUDA kernels.
    It then optimizes their runtime using a novel evolutionary meta-generation procedure tailored to the CUDA ecosystem, guided by LLM-based verifiers for correctness and efficient filtering.
    Evaluated on \texttt{robust-kbench}, our approach produces CUDA kernels outperforming \texttt{torch} implementations for practical applications, including forward and backward passes.
    It can fuse operations and deploy various runtime optimization strategies.
    The verifier workflow accurately classifies incorrect kernels, enhancing hardware verification efficiency.

\vspace{0.5em}
\begin{center}
\begin{tabular}{rcl}
\raisebox{-1.5pt}{\includegraphics[height=1.05em]{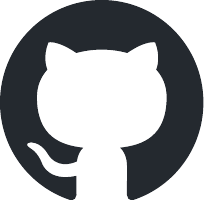}} & \textbf{Code} & \href{https://github.com/SakanaAI/robust-kbench}{\path{https://github.com/SakanaAI/robust-kbench}}\\
\end{tabular}
\end{center}
\end{abstract}

\section{Introduction}

The demand for computational power in machine learning has increased exponentially over the past decade, driven by the rising complexity of deep learning models and the need for large-scale data processing \citep{sevilla2022compute, hernandez2020measuring}.
The emergence of foundation models, which require incredible amounts of training and inference infrastructure \citep{jaech2024openai, claude2, claude3, liu2024deepseek, guo2025deepseek}, has exacerbated this trend and necessitates innovations in model efficiency and hardware acceleration \citep{bommasani2021opportunities}.
This has led to the proliferation of specialized hardware, such as GPUs, IPUs, and TPUs, optimized for deep learning workloads \citep{jouppi2017datacenter}.
However, the skill set required to balance trade-offs between hardware and software constraints and to engineer efficient CUDA kernels is both rare and highly sought after. It involves expertise in algorithms, hardware architecture, and instruction sets.
Simultaneously, there have been advancements in automated agentic discovery \citep[e.g.][]{romera2024mathematical, lu2024discovering, lu2024ai,Novikov2025AlphaEvolve} using Large Language Models (LLMs). A key advantage of LLM-guided discovery is the ability of these models to iteratively refine hypotheses, generate experiments, and interpret results, which facilitates closed-loop exploration \citep{yang2023large, song2024position}. 

\begin{figure}[ht]
    \centering
    \includegraphics[width=0.995\textwidth]{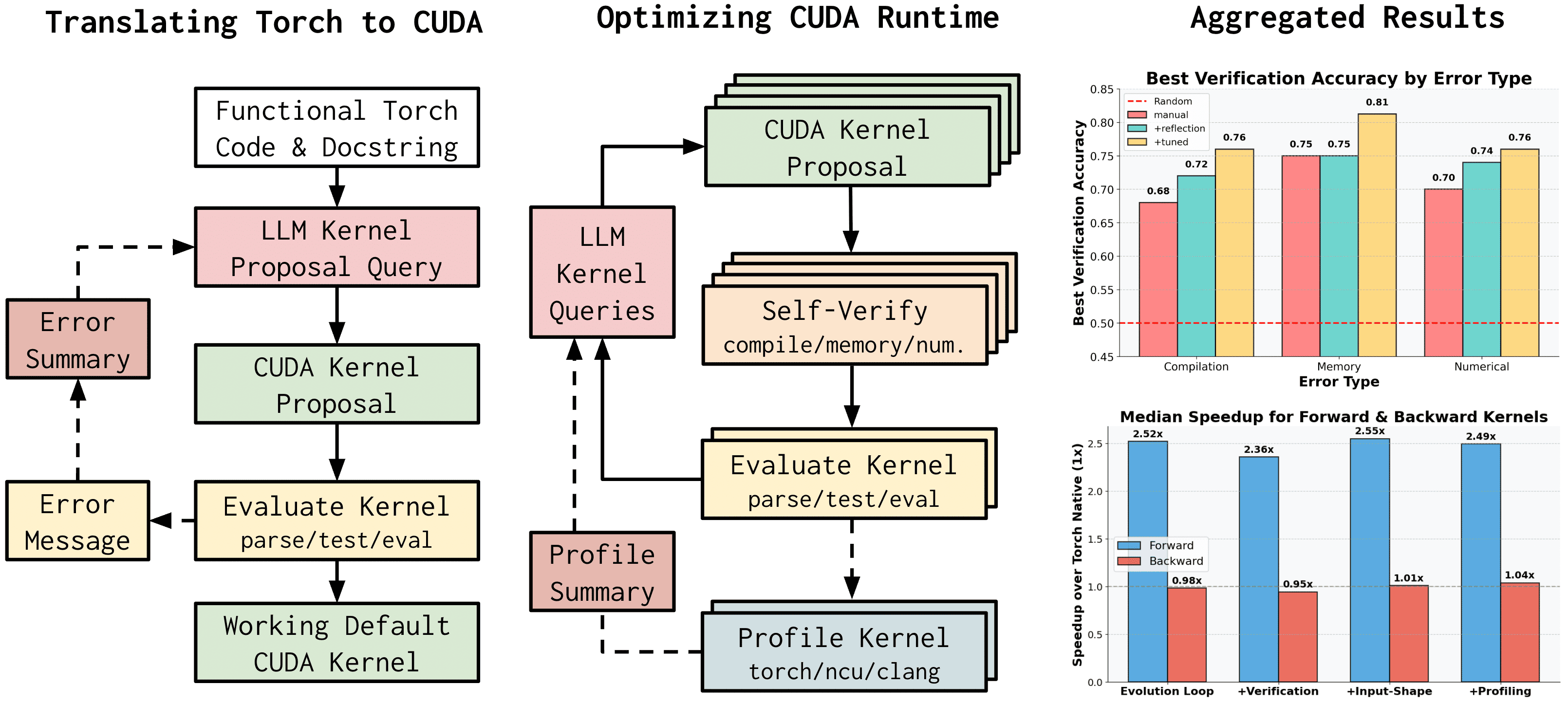}
    \caption{\textbf{High-level overview of the LLM-Driven CUDA Optimization \& Core Results.} \\ \underline{Left}: Functional PyTorch code is translated into a corresponding CUDA kernel, which is loaded to replace the PyTorch-eager operation. \underline{Middle}: We use the translated kernel to initialize a runtime optimization process, which samples, verifies, tests, and evaluates a batch of kernels in parallel. Throughout, we use a series of language model-based verifiers to ensure correctness and efficient filtering of candidate kernels. \underline{Right}: We demonstrate that our approach can accurately identify incorrect kernels (top) and discovers high-performing kernels (bottom) on the proposed \texttt{robust-kbench}. Runtime improvements are harder to achieve for backward than for forward kernel computations.}
    \label{fig:conceptual}
    \vspace{-0.5cm}
\end{figure}

Here, we set out to answer a simple question: \textbf{Can we leverage recent agentic advances to improve the robust discovery, self-verification, and optimization of low-level CUDA operations?}
More specifically, we introduce an LLM-driven evolutionary optimization framework to directly optimize the runtime of kernel operations. In doing so, we also find detrimental shortcomings in the current state of benchmarking LLM-written CUDA kernels: Due to the exploitable loopholes in the benchmark design, LLMs are capable of finding exploits such as omitting redundant operations, overfit input settings, and implementations that do not generalize to practical applications. We propose a robust benchmark suited to properly evaluate our optimization framework within. Additionally, we introduce a new verifier workflow, which is capable of accurately identifying incorrect kernels, improving the success rate of hardware verification. Our contributions are summarized as follows:

\begin{enumerate}
    \item \textbf{A Robust Benchmark Harness:} We highlight several pitfalls of LLM-driven CUDA kernel optimization, including the exploitation of bad task design and narrow verification leading to artificial speedup estimates. Consequently, we introduce a new benchmark harness, \texttt{robust-kbench}, which tests the proposal correctness for various settings, enables forward and backward kernel optimization, and is dedicated to realistic downstream applications.
    \item \textbf{Soft-Verification:} We introduce an LLM-based soft-verification workflow, capable of accurately classifying incorrect kernels for compilation, memory access, and numerical correctness (up to 80\% accuracy). Our approach improves the proposal success rate using downstream hardware verification (up to 30\% increase) and facilitates test-time scaling.
    \item \textbf{Agentic E2E Workflow:} We introduce an end-to-end agentic workflow capable of translating PyTorch code to working CUDA kernels, optimizing CUDA runtime, and automatically fusing multiple operations. It leverages principles from evolutionary optimization, model-ensembling, in-context improvement, and kernel profile summarization.
    \item \textbf{Pipeline Analysis:} We provide various ablations of our approach analyzing the consistency and performance impact of the pipeline, including LLM ensembling, an iterative profiling summarization feedback loop, and the downstream impact of soft-verification filtering.
    \item \textbf{Open-Sourcing the Benchmark \& Dataset:} Along with this paper, we release the benchmark and an accompanying dataset of the discovered kernels, their profiling data, and self-verification results enabling future work on supervised fine-tuning and Reinforcement Learning post-training of both kernel proposal and verification models.
\end{enumerate}

\newpage
\section{Background}

\textbf{Scaling Test-Time Compute}. Due to the increasing costs associated with training in-house models, scaling test-time compute has emerged as a strong paradigm to improve LLM outputs outside of training a new model from scratch. There are numerous options for improving samples drawn from an LLM, from the utilization of search techniques \citep{xie2023selfevaluation, zhang2024restmcts} to the incentivization of reasoning during post-training \citep{guo2025deepseek, team2025kimi}. We focus on techniques that scale with the number of samples drawn. Parallel samples can be aggregated via voting methods \citep{snell2024scaling, trad2024ensemble} or LLM debates \citep{du2023improving, wang2025mixtureofagents}. A related strategy is evolutionary test-time compute \citep{berman2025record}, which aggregates samples by mutating and recombining them. Sequential samples, meanwhile, can utilize multi-turn structures \citep{zhou2024archer} to improve and correct previous responses. Parametric verifiers \citep{luo2024improve, snell2024scaling} can be used to score drawn samples based on their final outcome or even intermediate steps, while non-parametric \citep{li2025combining, deepmind2024ai} verifiers range from also providing scalar scores to filtering out incorrect answers via some heuristic. Verification can be performed by the same model used for generating responses (often called 'self-verification') or by ensembles of different verifiers \citep{lifshitz2025multi, zhao2025sample}.
Here, we focus on drawing multiple sequential streams of samples and aggregating information across them via evolutionary optimization (by selecting previous samples to place in context and recombine) guided by self-verifiers, kernel accuracy, speed, and profiles.

\textbf{Evolutionary Code Optimization with LLMs}. One particular flavor of test-time compute is evolutionary code optimization: the usage, mutation, and recombination of previously generated code to produce new samples. This approach has previously been used to optimize reward and preference objectives \citep{lu2024discovering, ma2023eureka}, mathematical science code \citep{romera2024mathematical}, entire machine learning papers \citep{lu2024ai}, and other applications \citep{lehman2022evolutionlargemodels, lange2024large, meyerson2023language, berman2025record}. Through prompting, LLMs are used as recombination engines \citep{lange2023discovering_es, meyerson2023language}, and are capable of simulating crossover between diverse code snippets and the rationales that produced them. A simpler form of this technique is retrieval augmented generation \citep[RAG,][]{lewis2020retrieval, gao2023retrieval}, whereby historical samples are injected into context based on embedding similarities or other filters.

\textbf{The CUDA Framework}. CUDA is a parallel computing platform and application programming interface (API) developed by NVIDIA for leveraging the massive parallel processing power of GPUs. It extends the C and C++ programming languages with GPU-specific extensions, allowing developers to write efficient parallelized code for tasks such as deep learning, scientific computing, and real-time graphics rendering \citep{nickolls2008scalable}. CUDA enables fine-grained control over GPU threads, memory hierarchies, and synchronization mechanisms, making it an essential tool for accelerating workloads that require extensive matrix and tensor computations \citep{kirk2016programming}. Its core model consists of a hierarchy of threads organized into warps, blocks, and grids, facilitating scalable parallel execution \citep{garland2008parallel}. CUDA also provides libraries such as cuBLAS for linear algebra, cuDNN for deep learning, and Thrust for high-level parallel algorithms, further optimizing computational efficiency \citep{chetlur2014cudnn}. 

\textbf{Challenges for Benchmarking LLM-Written CUDA Kernels}. The KernelBench v0 \citep{ouyang2025kernelbenchllmswriteefficient} benchmark introduced a set of 250 neural network tasks, defined as PyTorch modules, along with a subset of corresponding results for CUDA kernel generation across these tasks. The tasks are split into three categories denoted as levels 1, 2, and 3. 
These tasks represent a natural gradation from producing high-quality operators, to fusing them together, to assembling them into a larger neural network. \citet{metr2025measuring} noted that approximately 40 of these tasks are problematic in various ways. First, many benchmark tasks include flaws such as inefficient PyTorch-eager baselines, low-magnitude outputs that can be dominated by precision errors, and insufficient output variation across different seeds, which enables easy exploitation, as we will demonstrate in this work. Second, run-time profiling of these tasks is often compromised by CPU overhead, which can dominate measurement for small kernels and lead to misleading optimization targets. KernelBench remains the de facto standard for benchmarking LLM-written CUDA kernels \citep{nvidia_deepseek_r1, multi-turn-kernels}.

\textbf{Challenges for LLM-Driven Optimization of CUDA Kernels}. LLM-generated CUDA kernels face several critical challenges that can undermine their practical utility. First, LLMs can exploit benchmark loopholes through various forms of ``cheating'', such as eliminating redundant operations, hardcoding for specific input patterns, or making assumptions about weights that do not generalize beyond individual test cases \citep{lange2025ai}. Second, many LLM-optimized kernels fail to translate their benchmark performance to real-world applications due to their inability to handle diverse input shapes, precision requirements, and integration with existing frameworks. These kernels often optimize for narrow test cases rather than the variable conditions encountered in production environments, resulting in solutions that appear impressive in isolation but prove impractical when deployed in actual machine learning pipelines. We aim to address these challenges through our new benchmark.

\newpage
\section{\texttt{robust-kbench}: A Robust Agentic CUDA Kernel Discovery Benchmark}

\begin{figure}[ht]
    \centering
    \includegraphics[width=0.995\textwidth]{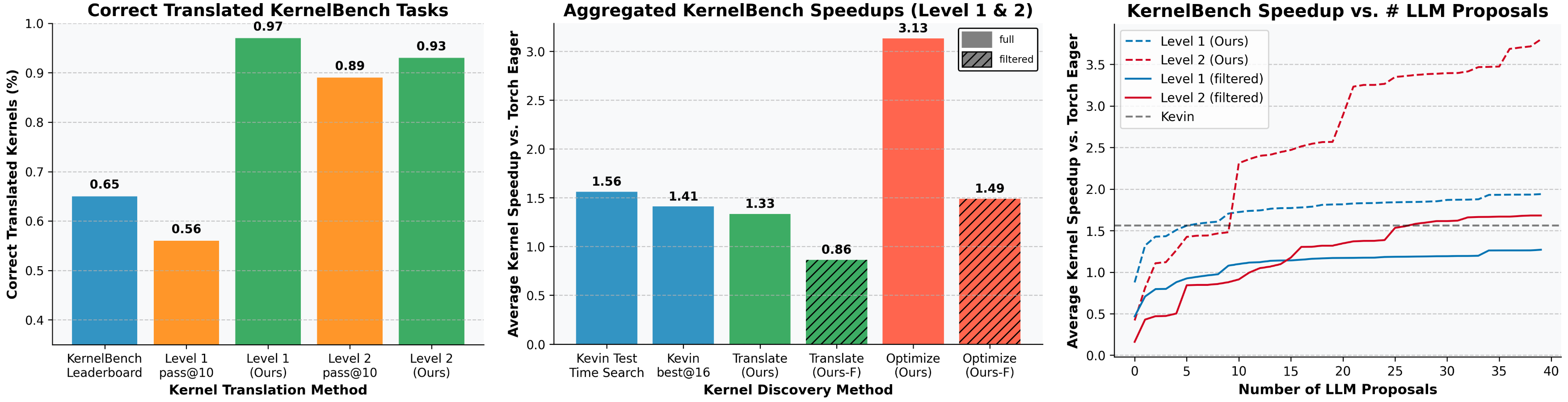}
    \caption{\textbf{KernelBench Tasks.} \underline{Left:} Our proposed translation approach successfully translates 95\% of all level 1 and level 2 KernelBench tasks. Incorporated LLM summarization of error messages outperforms simple parallel sampling. \underline{Middle:} Our proposed agentic optimization framework significantly outperforms the Kevin-32B model, both evaluated on the full 200 tasks. After excluding contaminated tasks, the aggregated speedup significantly reduces. \underline{Right:} Our evolutionary optimization approach displays test-time scaling behavior, discovering better speedups with more tries.}
    \label{fig:kbench}
\end{figure}

As outlined in the previous section, benchmarking LLM-written CUDA kernels presents significant challenges. To highlight the severity of these, we first revisit the 200 level 1 and level 2 KernelBench tasks. More specifically, we run our proposed agentic translation and evolutionary optimization framework on these tasks and compare the results with \citet{multi-turn-kernels}, which fine-tuned a dedicated QwQ-32B model using reinforcement learning. They report results on the full 200 tasks. We find that our approach is capable of significantly outperforming the reported results (\Cref{fig:kbench}, middle). Upon closer inspection, these speedups largely result from the exploitation of KernelBench's loopholes. More specifically, after excluding all contaminated tasks (\Cref{appsec:kbench}), the average speedup is reduced from 3.13x to 1.49x. Please refer to \Cref{appsec:kbench} for examples of '\textit{cheating}' kernels which pass KernelBench's verification process, and even achieve \textit{fake} speedups of \textbf{50-120x} by exploiting loopholes in KernelBench.
Additionally, existing KernelBench tasks are only tested and evaluated on a single input configuration, making them unsuitable for discovering general-purpose kernels.

To address these limitations, we introduce a comprehensive benchmarking harness - \texttt{robust-kbench} - that provides a more robust evaluation framework. The harness implements multiple layers of testing and validation, including diverse initialization states to prevent hardcoding, multiple runtime estimation strategies to ensure consistent performance measurement, and integration with various profiling tools. Specifically, we leverage PyTorch's built-in profiler for high-level metrics, Clang-tidy for static analysis, and NVIDIA's Compute Profiler (NCU) for detailed hardware-level insights. This multi-faceted approach helps identify potential optimizations while ensuring kernels maintain correctness across realistic execution contexts.
Additionally, a key contribution of \texttt{robust-kbench} is its ability to evaluate both forward and backward pass computations. This represents an advancement over existing frameworks, which focus only on forward pass operations.
To demonstrate the practical utility of our framework, we introduce several new benchmark tasks that target common deep learning workloads. These include kernels for MNIST CNN training, which tests fundamental convolution and pooling operations, ResNet-18 inference, which evaluates more complex residual architectures, and Transformer Llama inference, which addresses the specific challenges of attention-based models. As detailed in Table \ref{tab:benchmark_tasks}, each task supports multiple initialization states, input configurations, forward and backward pass computations, providing a more comprehensive assessment of kernel robustness and efficiency.
To facilitate the evaluation of CUDA kernels, we provide a simple Python API that enables consistent testing across different tasks. Below is a minimal example of our evaluation interface:

\vspace{-0.25cm}
\begin{minipage}[t]{0.42\textwidth}
\begin{lstlisting}[caption=Benchmark Task Directory]
tasks                    # Base dir
|- mnist_cross_entropy   # Task dir
|-- func_forward.py      # Forward
|-- func_backward.py     # Autograd
|-- config_forward.json  # F. config
|-- config_backward.json # B. config
|-- forward.cu           # Op. kernel
|-- backward.cu          # Op. kernel
\end{lstlisting}
\end{minipage}
\hfill
\begin{minipage}[t]{0.54\textwidth}
\begin{lstlisting}[caption=Kernel Evaluation Python API]
from robust_kbench import ParallelKernelExecutor
executor = ParallelKernelExecutor(
    task_dir="tasks/mnist_cross_entropy",
    **task_specific_settings)
kernels = ["kernel_1.cu", "kernel_2.cu"]
torch_results = executor.torch_eval()
test_results = executor.test(kernels)
eval_results = executor.evaluate(kernels)
\end{lstlisting}
\end{minipage}

The task specification includes parameters such as input shapes, initialization settings, and whether to test forward and backward passes. The evaluator handles the complexity of kernel compilation, correctness verification, and profiling. Furthermore, it enables parallelization across multiple GPUs. We provide a detailed list of tasks and an example of a task specification in Appendix \ref{appsec:benchmark}.

\section{Automating CUDA Kernel Correctness Verification with LLMs}

The verification of CUDA kernel proposals presents scaling challenges in terms of computational resources and time efficiency. Traditional hardware-based verification requires compilation times of at least one minute per kernel, with parallelization constrained by the available GPU hardware. This bottleneck becomes particularly problematic when evaluating multiple kernel variants or conducting extensive optimization searches. Moreover, as discussed earlier, kernel proposals can potentially exploit benchmark tasks, necessitating robust verification methods that can detect such manipulations.

\begin{figure}[ht]
    \centering
    \includegraphics[width=0.995\textwidth]{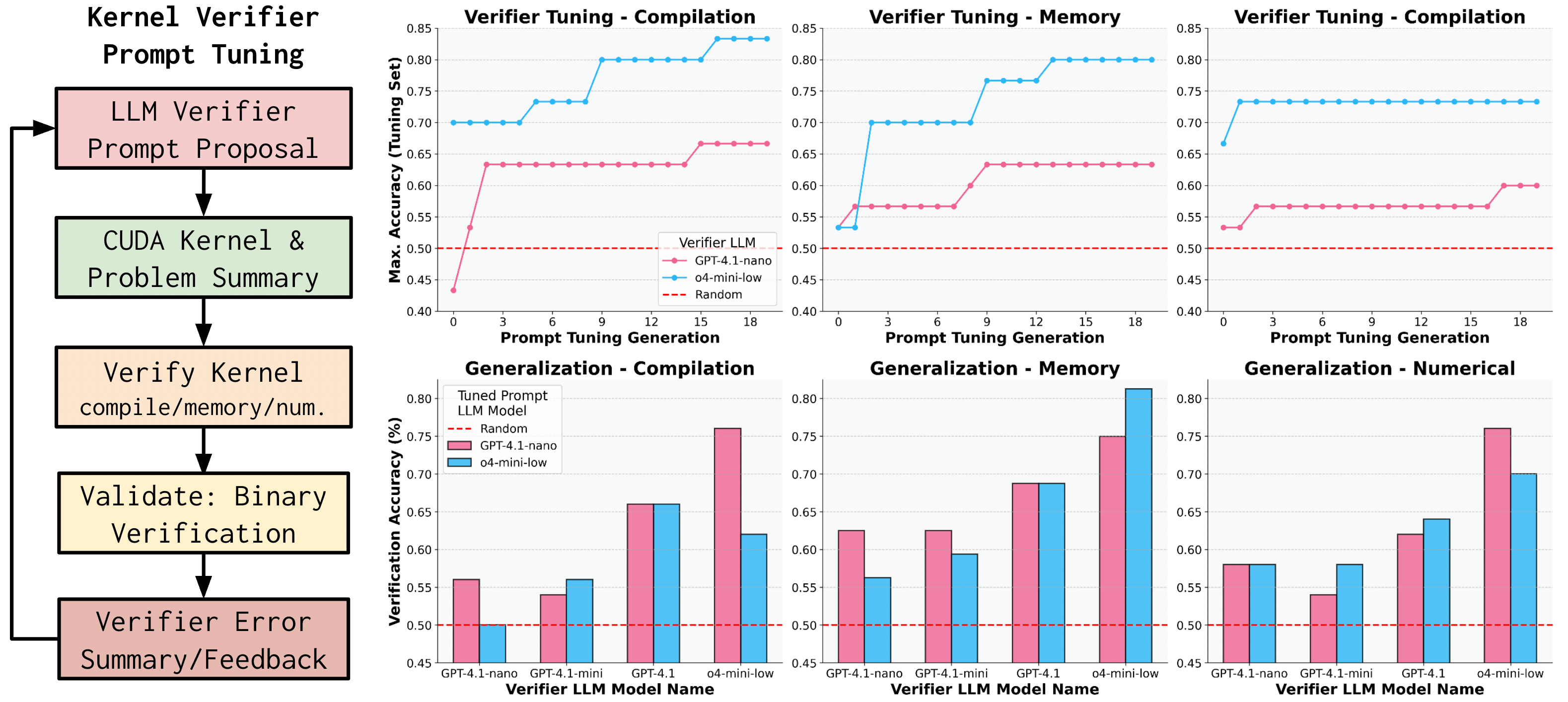}
    \caption{\textbf{Verifier Prompt Tuning Pipeline \& Results.} \underline{Left}: Overview of the LLM-based verifier prompt tuning workflow, where a dataset of kernel proposals is used to iteratively improve the LLM-based verifier's ability to detect errors. \underline{Right, Top}: Accuracy results across generations for specialized verifiers targeting different types of CUDA errors: compilation, memory, and numerics. \underline{Right, Bottom}: The tuned prompts generalize to different downstream verifier models.}
    \label{fig:verifier_tune}
\end{figure}

To address these challenges, we introduce an LLM-based verification workflow that significantly improves the effectiveness of the hardware verification process. Our approach leverages language models to perform rapid ``soft verification'' of kernel proposals before proceeding with hardware testing. For every kernel proposal, each LLM verifier makes a binary decision regarding its correctness.
The highest-scoring kernels (determined by majority consensus) proceed to hardware verification. LLM-based verification can easily be parallelized across queries and API endpoints, offering substantial speedups compared to traditional verification methods.
We found that manual construction of verification prompts is challenging (Figure \ref{fig:conceptual}, right) and used self-reflection \citep{shinn2024reflexion} to improve the robustness of the verifiers. Furthermore, we employed an iterative prompt tuning methodology to enhance the accuracy of our verifiers (Figure \ref{fig:verifier_tune}, left). We constructed datasets containing both correct and incorrect kernel implementations to train our verifiers. The training process involved a meta-agent that designs both system and message prompts, optimizing the verifier's ability to detect various types of errors in kernel proposals. We provide error summaries to continuously refine the verifier design.

We tuned three specialized verifiers targeting distinct aspects of CUDA kernel correctness: compilation, memory access, and numerical accuracy. Each verifier was tuned on a balanced dataset of 30 kernels. The results demonstrate the effectiveness of our approach, with the compilation verifier achieving an accuracy of 0.82, the memory access verifier reaching 0.80, and the numerical correctness verifier attaining 0.73 (\Cref{fig:verifier_tune}, top right).
Furthermore, the tuned prompts (\Cref{appsec:verifier_prompts}) demonstrate strong generalization capabilities, successfully evaluating $\sim$ 20 previously unseen kernels and maintaining their performance across different LLM base models (\Cref{fig:verifier_tune}, bottom right). Next, we investigate the impact of our verification process on the kernel optimization process.

\section{Automating CUDA Kernel Discovery with LLMs}
\vspace{-0.4cm}
\textbf{Kernel Translation \& Optimization}. We translate PyTorch code into a working CUDA kernel by querying an LLM with the corresponding functional \texttt{torch} implementation. After parsing the LLM's output, we load the kernel using the \texttt{torch} C++ loading utilities. We evaluate the kernel's numerical correctness by comparing its results with the \texttt{torch} reference implementation. If the compilation fails or the computed result does not lead to a correct computation (breaching $1e-5$ precision), we summarize the error message using an additional LLM call (\Cref{fig:conceptual}, left). This information is fed back to the LLM before we iterate. We found that this procedure improves the robustness of correct translations compared to best-of-N sampling \citep{brown2024large} with the same compute budget (\Cref{fig:kbench}, left).

\input{evo_algo.tex}

We use the working CUDA kernel from translation to initialize an evolutionary optimization process which samples, self-verifies, evaluates, and profiles batches of kernels to improve the runtime (\Cref{fig:conceptual}, middle, \Cref{alg:evo_cuda}). Given a set of previous kernel evaluations, we filter them for correctness and provide a subset of up to five correct kernels sorted from slowest to fastest \citep{zhou2022least}. This least-to-most ordering incentivizes the LLM to infer optimization patterns from simpler to more sophisticated implementations.
We sample $N=8$ proposals from an LLM ensemble including both reasoning (o3 \& o4-mini) and conventional LLMs (Claude Sonnet 3.7 \& GPT-4.1). Additionally, we perform temperature sampling when applicable \citep[][, see \Cref{appsec:hyperparams}]{renze2024effect}. Inspired by AlphaCode's \citep{li2022competition} prompting approach, we sample high-level recommendations (e.g., optimizing block sizes, using stride loops, etc.) to encourage diversity in model outputs. Afterwards, we perform the verification filtering step and evaluate $N^\star=4$ kernels on the hardware accelerator. For each correct kernel, we obtain the \texttt{torch}, \texttt{NCU}, and \texttt{Clang-tidy} profiling information. Additionally, we experiment with providing input shape information during prompting and LLM summaries of the profiling data. All kernels are evaluated on individual H100 GPUs using CUDA 12.4 (\Cref{appsec:eval_env}).

\begin{figure}[htbp]
    \centering
    \includegraphics[width=0.995\textwidth]{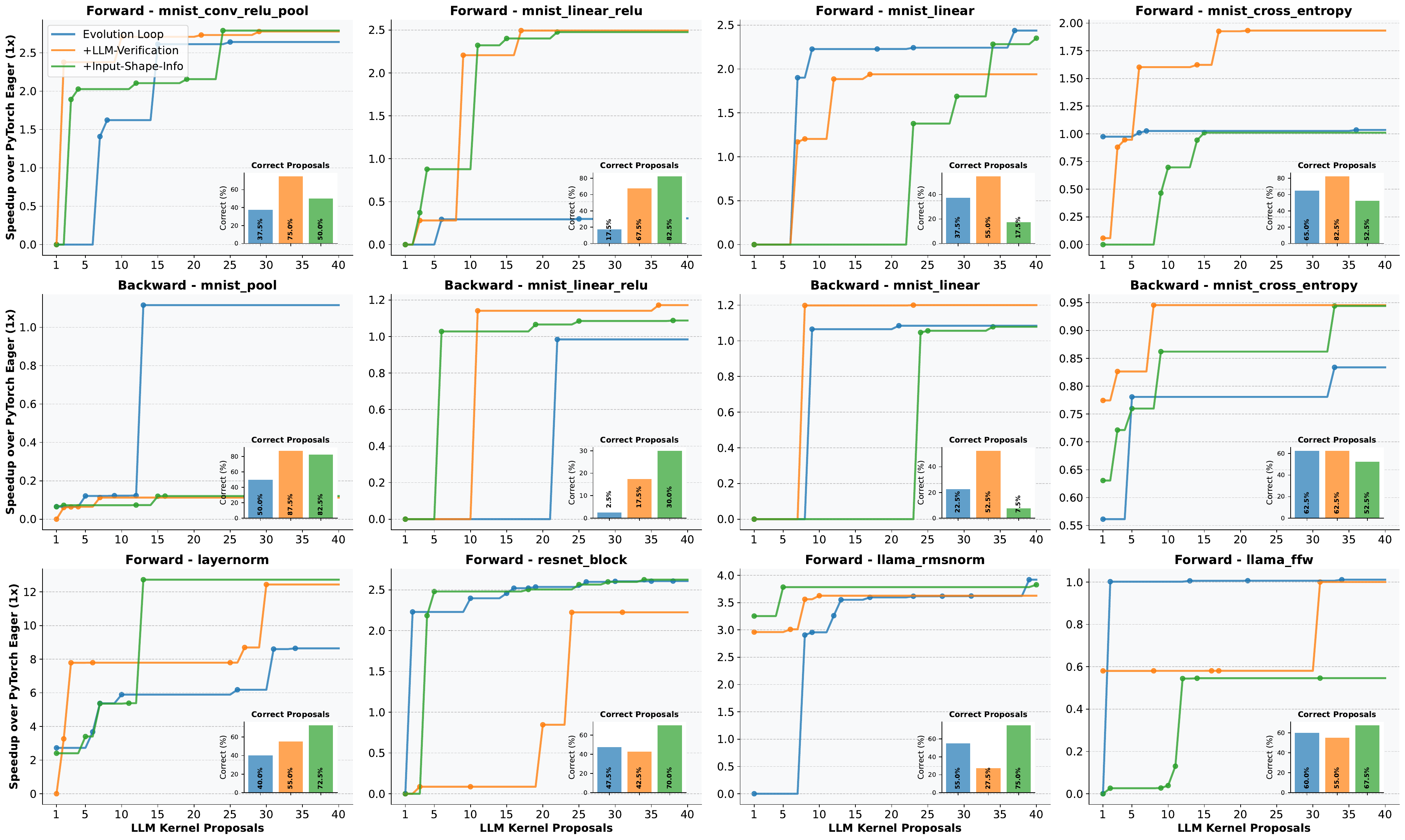}
    \caption{\textbf{Forward and Backward Pass Speedups.} Speedup of LLM-optimized CUDA kernels, with and without verifier, and input shape information on twelve tasks. Forward pass achieves up to 2.5x speedup; backward pass yields smaller but consistent gains. The verifier improves stability and successful kernel evaluation (yellow). Adding input shape information to the system prompt can improve performance (green). Improvements scale with the number of kernel proposals.}
    \label{fig:optimize}
\end{figure}

\textbf{Results: Optimizing individual operations}. The results in Figure~\ref{fig:optimize} demonstrate the effectiveness of our evolutionary optimization framework for CUDA kernels. Across multiple \texttt{robust-kbench} tasks, the evolutionary loop rapidly discovers kernel proposals that outperform the PyTorch eager baseline, achieving up to 2.5$\times$ speedup in the forward pass. The inclusion of LLM-based verification further improves the stability of the optimization process, reducing the incidence of regressions and increasing the proportion of correct kernel proposals. Additional feedback mechanisms, such as providing input shape in the system prompt, can yield further performance gains. Notably, the backward pass also benefits from these strategies, though the speedup there is more modest compared to the forward pass. In general, we find that optimization of backward kernels is significantly more challenging. We hypothesize that this could be potentially due to the unavailability of pre-training data and the increased complexity of caused by combining kernel fusion and activation recomputation. 

\begin{figure}[htbp]
    \centering
    \includegraphics[width=0.995\textwidth]{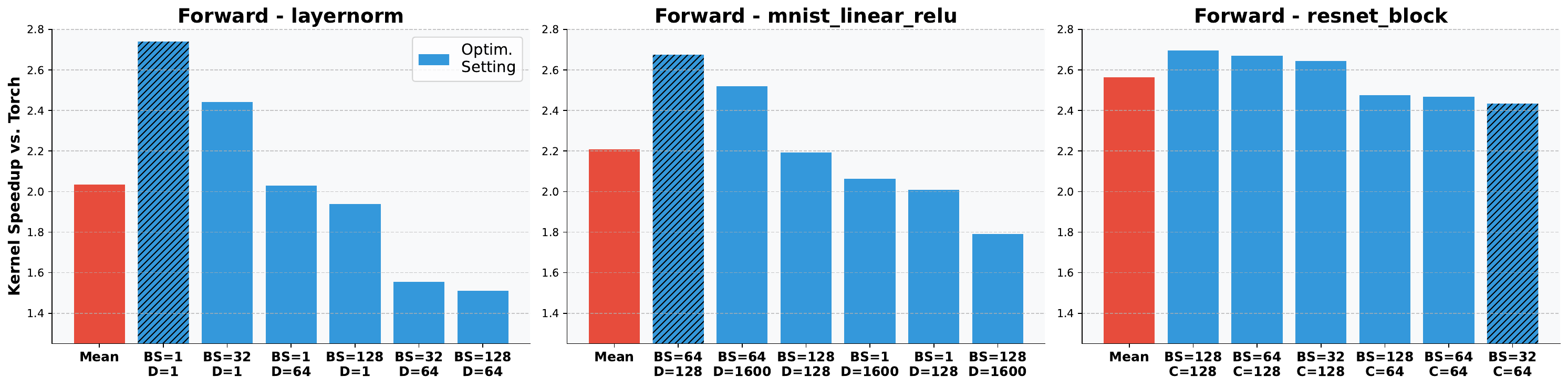}
    \caption{\textbf{Generalization of discovered kernels to unseen input shapes.} We evaluate the optimized CUDA kernels on input shapes not seen during optimization. For LayerNorm and MNIST Linear-ReLU tasks, the kernels show signs of overfitting to the training configuration, with performance degrading on unseen shapes. For the ResNet block task we observe positive generalization, with the optimized kernels maintaining their performance benefits across different input dimensions.}
    \label{fig:generalization}
\end{figure}

\textbf{Results: Generalization of LLM-Optimized Kernels}. While our correctness verification considers multiple kernel settings, the previous results are achieved by optimizing the runtime for a single configuration. To assess the robustness of the optimized kernels, we evaluate their performance on input shapes not encountered during the evolutionary optimization process. Unlike KernelBench, which only tests kernels on fixed input configurations, \texttt{robust-kbench} enables systematic evaluation of kernel generalization across diverse input shapes. As shown in Figure~\ref{fig:generalization}, we observe varying degrees of generalization across different tasks. For simpler operations like LayerNorm and MNIST Linear-ReLU, the optimized kernels tend to overfit to the specific input shapes used during training, with performance degrading when tested on unseen configurations. This suggests that the optimization process may leverage task-specific patterns that don't transfer well to different input dimensions. However, for the more complex ResNet task, we find that the optimized kernels maintain their performance benefits across different input shapes. These findings highlight the importance of considering generalization properties when evaluating kernel optimization approaches, as the ability to maintain performance across different input configurations is crucial for practical deployment.

\newpage
\section{Ablating the Agentic Scaffolding for Kernel Verification \& Optimization}

\begin{figure}[htbp]
    \centering
    \includegraphics[width=0.975\textwidth]{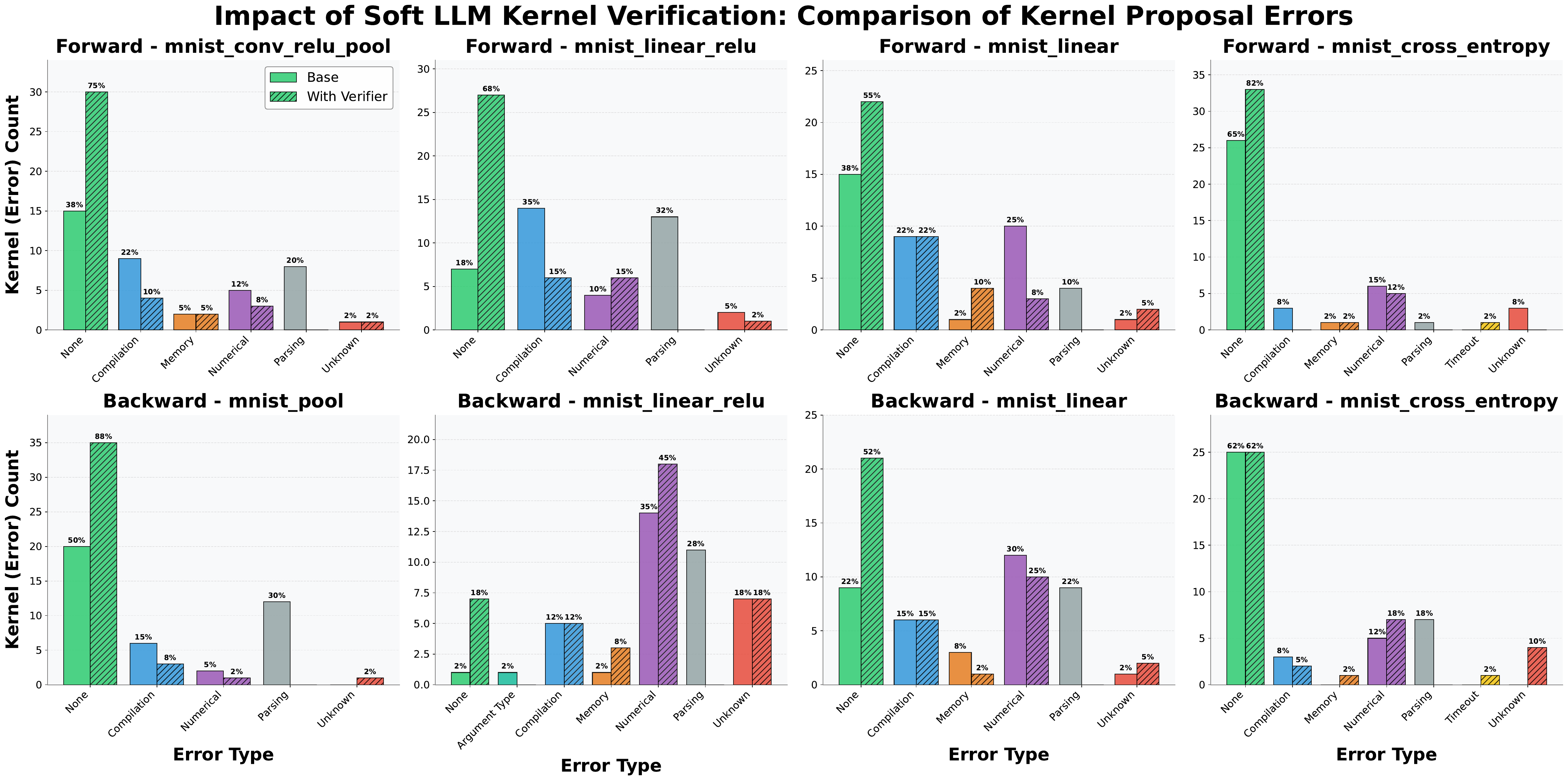}
    \caption{\textbf{Error Type Distribution Impact of Verifier.} Solid bars show base optimization without verification; hatched bars show verifier-assisted optimization. This pre-screening mechanism effectively filters out problematic kernels before hardware evaluation, improving efficiency.}
    \label{fig:verifier_impact}
\end{figure}

\textbf{Impact of LLM Verification}. To evaluate our LLM-based verifier, we analyzed its performance across eight CUDA operations, including both forward and backward passes. The verifier was tested on its ability to detect three types of errors: compilation failures, memory access violations, and numerical inaccuracies. Our results (\Cref{fig:verifier_impact}) show that verifier-assisted optimization significantly outperforms the baseline approach, increasing the proportion of valid kernels from 55-70\% to 80-85\% in forward passes. The improvement was particularly effective at filtering out compilation errors, which are prevalent in forward pass implementations. Additionally, the kernel code parsing was improved likely due to an increased number of in-context examples. This pre-screening mechanism reduced computational overhead by preventing invalid kernels from reaching hardware testing.

\begin{figure}[htbp]
    \centering
    \includegraphics[width=0.975\textwidth]{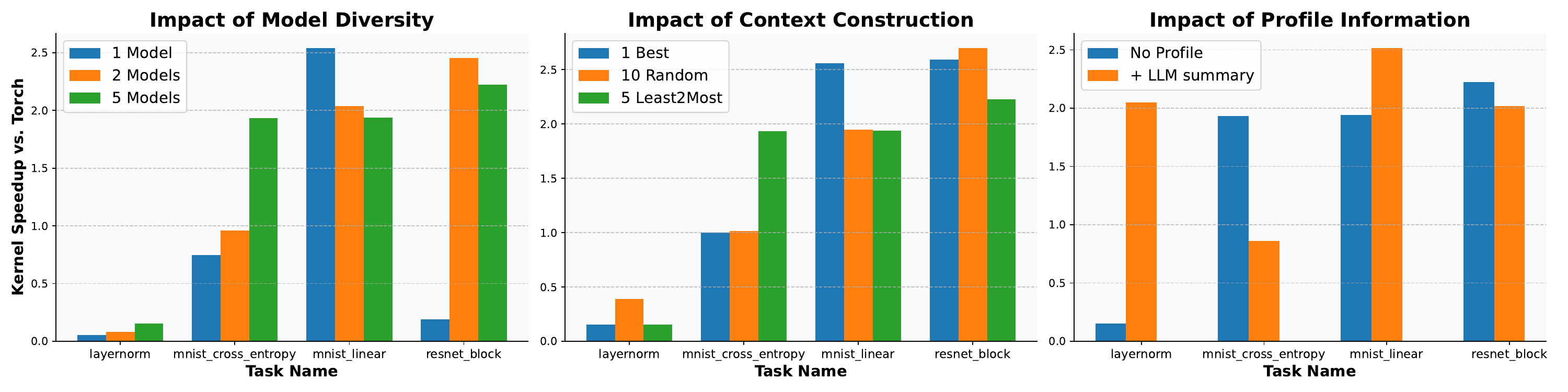}
    \caption{\textbf{Optimization Framework Ablations Across 4 Tasks.} \underline{Left}: Impact of model ensembling. \underline{Middle}: Effect of context construction strategies. \underline{Right}: Contribution of profiling feedback.}
    \label{fig:ablations}
\end{figure}

\textbf{Impact of Model Ensembling}. We analyze the effect of model ensembling on optimization performance, as depicted in the left panel of \Cref{fig:ablations}. We compare using a single model (GPT-4.1), a two-model ensemble (GPT-4.1 \& Claude Sonnet 3.7), and a five-model ensemble (GPT-4.1, o3, o4-mini, Claude Sonnet 3.7, and Gemini 2.5 Pro). The results indicate a positive trend, with increasing diversity in the model ensemble leading to improved optimization outcomes and a higher success rate in generating valid and performant kernels.\\
\textbf{Impact of Context Construction}. \Cref{fig:ablations} (middle) illustrates the impact of different context construction strategies for prompting the LLM during the optimization process. We evaluated three approaches: providing only the single best-performing kernel, providing the top five best-performing kernels in a least-to-most sorted order, and providing ten randomly selected previous kernels. The least-to-most strategy tends to yield the most consistent improvements, suggesting that a curated and ordered set of examples aids the model in identifying effective patterns.

\textbf{Impact of Additional Profiling Information}. We investigated the utility of providing summarized profiling information as feedback to the optimization pipeline, with results shown in the right panel of \Cref{fig:ablations}. One setup involved no explicit profiling feedback, while the other provided LLM-summarized insights from \texttt{torch}, \texttt{NCU}, and \texttt{Clang-tidy} profilers. Incorporating summarized profiling data leads to more targeted kernel modifications and measurably better performance, highlighting the value of detailed, yet digestible, hardware-level feedback in guiding the search.

\vspace{-0.2cm}
\section{Related Work}
\vspace{-0.2cm}

\textbf{Modern GPU Programming}. Recent advances in GPU programming frameworks have fostered an ecosystem that emphasizes both high performance and developer productivity. Frameworks like Triton \citep{tillet2019triton} offer a Python-based interface that allows developers to write efficient, low-level GPU kernels without delving into the intricate details of CUDA or OpenCL. Emerging tools such as ThunderKittens \citep{spector2024thunderkittens} focus on enhanced human usability and seamless integration with popular machine learning libraries, further simplifying complex workflows.\\
\textbf{Scientific Discovery with LLMs}. As LLMs become more capable, interest has grown in using them to aid scientific discovery. Prior works have shown the effectiveness of LLMs for paper review \citep{zhuang2025large, liu2023reviewergpt}, idea generation \citep{si2024can, dasgupta2024empowering, 10.1145/3715964}, and research synthesis \citep{agarwal2024litllm, ali2024automated}. Other works have gone further in integrating LLMs across the scientific process, having them propose, implement, and evaluate preference optimization objectives \citep{lu2024discovering} as well as entire scientific experiments \citep{lu2024ai, yamada2025ai}. The latter work is additionally capable of visualizing its results, writing a paper, and preparing it for review.\\
\textbf{Software Engineering with LLMs}. Another domain LLMs have shown promising results in is software engineering. Common benchmarks test standalone programming questions \citep{austin2021program, quan2025codeelo, metr2025measuring} as well as traditional software engineering tasks, such as completing GitHub issues \citep{jimenez2024swebench}.
Different methods diverge in how they tackle these applications. Some works train LLMs to become better coders \citep{zhu2024deepseek, hui2024qwen2, roziere2023code}, while others leverage test-time compute strategies to improve model outputs. There is also variation in the output form factor, which ranges from full files to iterative diffs \citep{aider}.\\
\textbf{Automated CUDA Kernel Discovery with LLMs}. A growing subset of work has begun to target kernel writing \citep{ouyang2025kernelbenchllmswriteefficient, zhang2025adaptive} in CUDA, since efficiency improvements from these kernels can be incredibly valuable and skilled human kernel engineers are in high demand. \citet{metr2025measuring} leverage LLMs for CUDA kernel generation. \citet{nvidia_deepseek_r1} report promising results for optimizing attention kernels using DeepSeek-R1. Unlike our approach, these approaches do not detail an end-to-end system that combines evolutionary test-time scaling, self-verification, model ensembling, or profiling data.

\vspace{-0.2cm}
\section{Discussion}
\vspace{-0.2cm}

\textbf{Summary}. We demonstrated that current LLM-written kernel benchmarks often lead to artificial speedups, obfuscating critical performance assessment. Consequently, we introduced \texttt{robust-kbench}, which combats these shortcomings. Additionally, we introduced a framework for automatic CUDA kernel discovery, self-verification and optimization. Our approach demonstrates that LLMs, when combined with evolutionary optimization, soft-verification, and hardware profiling, can successfully translate and optimize PyTorch operations into efficient CUDA implementations.\\
\textbf{Costs \& Runtime}. Our optimization results were produced with an estimated per-kernel cost of \$5 in API credits for foundation model usage and less than 2 hours of runtime on 4 GPUs including kernel generation, verification, and profiling (see \Cref{appsec:costs}). Costs and runtime can be scaled to increase translation and optimization effectiveness as well as cover more complex tasks.\\
\textbf{Future Directions \& Limitations}. While this work demonstrates the discovery of inference kernel improvements targeting specific settings, obtaining similar results for backward kernels remains difficult. In general, obtaining ''free lunch'' improvements across the spectrum of input shapes remains a challenge. With this project, we release the benchmark and discovered kernels, speedup times, profiling and error messages. We hope that this can aid to post-train open-source models via supervised fine-tuning or off-policy reinforcement learning (RL) as in \citep{guo2025deepseek,multi-turn-kernels}. 
Additionally, there are several improvements to be made with regard to targeting specific hardware -- producing kernels that utilize the specific strengths and instruction details.\\
\textbf{Broader Impact Statement}. Our work aims to democratize high-performance computing by making GPU programming more accessible to developers without specialized expertise. While this could reduce computational costs and energy usage through more efficient kernels, we acknowledge that it may widen the resource gap between organizations with and without access to powerful LLMs and GPUs. We encourage users to carefully consider the downstream environmental impact.

\newpage
\bibliographystyle{plainnat}
\bibliography{references}

\newpage
\appendix

\section*{\LARGE Supplementary Material - NeurIPS 2025 Submission}
\section*{Towards Robust Agentic CUDA Kernel Benchmarking, Verification, and Optimization}

\vspace*{20pt}
\section*{Table of Contents}
\vspace*{-5pt}
\startcontents[sections]
\printcontents[sections]{l}{1}{\setcounter{tocdepth}{2}}

\clearpage

\input{appendix.tex}



\end{document}

%% file: evo_algo.tex
\begin{algorithm}[H]
    \caption{Evolutionary Kernel Optimization with Verification \& In-Context Improvement}
    \label{alg:evo_cuda}
    \SetAlgoLined
    \KwIn{Initial Kernel $(K^\mathrm{init}, p^\mathrm{init})$, Generations $G$, Population Size $N$, Effective Pop. Size $N^\star$}
    \KwOut{Optimized LLM-Written CUDA Kernel $K^*$}
    Initialize archive $A = \{(K^\mathrm{init}, p^\mathrm{init})\}$ \tcp*{Initial few-shot examples}
    Set self-verifier model prompts $V = \{V_\mathrm{comp}, V_\mathrm{mem}, V_\mathrm{num}\}$\;
    \For{$g = 1$ to $G$}{
        \tcp{Sample LLM ensemble, context \& Generate kernel samples, self-verify} 
        $S = \{\}$\;
        \For{$i=1$ to $N$} {
            $\theta^g_i, C^g_i \gets \KwSty{SampleSettingsPlusContext}(A)$\;
            $K^g_i \gets \KwSty{SampleKernelWithLLM}(C^g_i, \theta^g_i)$\;
            $s_i \gets V_\mathrm{comp}(K^g_i) + V_\mathrm{mem}(K^g_i) + V_\mathrm{num}(K^g_i)$\;
            $S \gets S \cup \{s_i\} $\;
        }
        
        \tcp{Hardware parallel verification, evaluation \& profiling}
        \For{$i \in \KwSty{TopIndices}(S, N^\star)$} {
            $p_i \gets \KwSty{TestEvalProfileOnGPU}(K^g_i)$ \tcp*{$p_i$ contains performance \& profile}
            $A \gets A \cup \{(K^g_i, p_i)\}$ \tcp*{Update kernel archive}
        } 
    }
    \Return $K^* \gets \KwSty{SelectBestKernel}(A)$\;
\end{algorithm}

%% file: appendix.tex
\section{Compromised KernelBench Tasks}
\label{appsec:kbench}

\subsection{Filtering Procedure}

As discussed in the main text, existing benchmarks like KernelBench can suffer from several vulnerabilities that allow for artificial performance gains.
Our analysis is based on \href{https://github.com/ScalingIntelligence/KernelBench/tree/7dd9cfa0c02e8340c9bc9a919d3f19777a8f2eb2}{KernelBench v0 (commit 7dd9cfa)}. We filter tasks from the first two levels of the KernelBench dataset based on a comprehensive set of criteria designed to ensure the quality and relevance of the benchmark.
These "contaminated" tasks may exhibit issues such as inefficient baseline implementations, low-magnitude outputs where floating-point inaccuracies can obscure true computational correctness, or insufficient output variation across different seeds, which allows for trivial solutions or overfitting to specific test conditions.
LLMs can exploit these loopholes by, for example, removing ostensibly redundant operations (that may be critical for generality) or hardcoding solutions for specific input patterns.
Such exploitations lead to kernels that show impressive speedups on the benchmark but fail to generalize to real-world applications with diverse input shapes, precision requirements, and integration needs.
Our filtering procedure, detailed below, is designed to mitigate these issues and identify tasks that are truly representative of practical kernel optimization challenges.

Our filtering criteria are inspired by \citet{metr2025measuring}. Tasks are excluded if their outputs fall within a narrow range of -0.01 to 0.01, as this indicates a low signal-to-noise ratio where floating point inaccuracies can dominate computational correctness.
Similarly, we remove tasks exhibiting insufficient output variation, specifically those with an output standard deviation of less than 0.01 across different model and input seeds.
Another criterion targets tasks with overly uniform output tensors across their axes, as such uniformity presents an unrealistically simplified optimization challenge.
Furthermore, tasks are filtered out if they demonstrate minimal input impact, defined as an output variation of less than 0.01 across various random input seeds.
Finally, we leverage Sonnet-3.7 to identify and exclude tasks where the baseline operation is inherently inefficient or involves redundant computations that do not affect the final output.
These combined filtering steps aim to curate a robust set of tasks that are both challenging and representative of real-world kernel optimization scenarios, addressing the contamination issues discussed in the main text (see Section 3). We provide examples of two 'cheating' kernels on compromised tasks in \Cref{appsec:kbench_examples_1,appsec:kbench_examples_2}.

All filtering results can be viewed in detail here:

\begin{center}
\begin{tabular}{rcl}
\raisebox{-1.5pt}{\includegraphics[height=1.05em]{figures/github-logo.pdf}} & \textbf{Code} & \href{https://github.com/SakanaAI/robust-kbench}{\path{https://github.com/SakanaAI/robust-kbench}} \\
\end{tabular}
\end{center}

\subsection{Compromised Level 1 Tasks}
\begin{table}[H]
\centering
\footnotesize
\begin{tabular}{cc|c|ccccc}
\hline
\textbf{Level} & \textbf{Task} & \textbf{Task Name} & \makecell{\textbf{Output}\\\textbf{Range}} & \makecell{\textbf{Output}\\\textbf{Std}} & \makecell{\textbf{Output}\\\textbf{Axes}} & \makecell{\textbf{Input}\\\textbf{Impact}} & \makecell{\textbf{Baseline}\\\textbf{Inefficient}} \\
\hline
1 & 12 & Matmul\_with\_diagonal\_matric & False & False & False & False & True \\
1 & 13 & Matmul\_for\_symmetric\_matric & False & False & False & False & True \\
1 & 14 & Matmul\_for\_upper\_triangular & False & False & False & False & True \\
1 & 15 & Matmul\_for\_lower\_triangular & False & False & False & False & True \\
1 & 18 & Matmul\_with\_transposed\_both & False & False & False & False & True \\
1 & 23 & Softmax & True & True & True & True & False \\
1 & 36 & RMSNorm\_ & False & False & False & False & True \\
1 & 37 & FrobeniusNorm\_ & True & True & True & True & False \\
1 & 38 & L1Norm\_ & True & True & True & True & False \\
1 & 39 & L2Norm\_ & False & False & True & False & False \\
1 & 50 & Product\_reduction\_over\_a\_d & True & True & True & True & False \\
1 & 69 & conv\_transposed\_2D\_\_asymme & False & False & False & False & True \\
1 & 88 & MinGPTNewGelu & False & False & False & False & True \\
1 & 91 & cumsum\_reverse & False & False & False & False & True \\
1 & 92 & cumsum\_exclusive & False & False & False & False & True \\
1 & 94 & MSELoss & False & True & True & True & False \\
1 & 96 & HuberLoss & False & True & True & True & False \\
1 & 97 & CosineSimilarityLoss & False & True & True & True & False \\
1 & 98 & KLDivLoss & False & True & True & True & True \\
\hline
\end{tabular}
\caption{KernelBench Tasks with Compromised Properties (Level 1)}
\label{tab:kbench_1}
\end{table}

\subsection{Compromised Level 2 Tasks}

\begin{table}[H]
\centering
\scriptsize
\begin{tabular}{cc|c|ccccc}
\hline
\textbf{Level} & \textbf{Task} & \textbf{Task Name} & \makecell{\textbf{Output}\\\textbf{Range}} & \makecell{\textbf{Output}\\\textbf{Std}} & \makecell{\textbf{Output}\\\textbf{Axes}} & \makecell{\textbf{Input}\\\textbf{Impact}} & \makecell{\textbf{Baseline}\\\textbf{Inefficient}} \\
\hline
2 & 2 & ConvTranspose2d\_BiasAdd\_Clam & False & False & False & False & True \\
2 & 4 & Conv2d\_Mish\_Mish & False & False & False & False & True \\
2 & 6 & Conv3d\_Softmax\_MaxPool\_MaxP & False & False & False & False & True \\
2 & 7 & Conv3d\_ReLU\_LeakyReLU\_GELU\ & False & False & False & False & True \\
2 & 8 & Conv3d\_Divide\_Max\_GlobalAvg & False & False & False & False & True \\
2 & 9 & Matmul\_Subtract\_Multiply\_Re & True & True & True & True & True \\
2 & 10 & ConvTranspose2d\_MaxPool\_Hard & False & False & False & False & True \\
2 & 12 & Gemm\_Multiply\_LeakyReLU & False & False & False & False & True \\
2 & 13 & ConvTranspose3d\_Mean\_Add\_So & False & True & True & True & True \\
2 & 14 & Gemm\_Divide\_Sum\_Scaling & False & False & False & False & True \\
2 & 15 & ConvTranspose3d\_BatchNorm\_Su & False & False & False & False & True \\
2 & 18 & Matmul\_Sum\_Max\_AvgPool\_Log & False & False & False & False & True \\
2 & 20 & ConvTranspose3d\_Sum\_Residual & False & False & False & False & True \\
2 & 22 & Matmul\_Scale\_ResidualAdd\_Cl & False & False & False & False & True \\
2 & 23 & Conv3d\_GroupNorm\_Mean & False & True & True & True & False \\
2 & 25 & Conv2d\_Min\_Tanh\_Tanh & False & False & False & False & True \\
2 & 26 & ConvTranspose3d\_Add\_HardSwis & False & False & False & False & True \\
2 & 27 & Conv3d\_HardSwish\_ReLU\_Softm & False & True & True & True & True \\
2 & 28 & BMM\_InstanceNorm\_Sum\_Residu & False & False & False & False & True \\
2 & 29 & Matmul\_Mish\_Mish & False & False & False & False & True \\
2 & 31 & Conv2d\_Min\_Add\_Multiply & False & False & False & False & True \\
2 & 33 & Gemm\_Scale\_BatchNorm & False & False & False & False & True \\
2 & 34 & ConvTranspose3d\_LayerNorm\_GE & False & False & False & False & True \\
2 & 36 & ConvTranspose2d\_Min\_Sum\_GEL & False & False & True & True & False \\
2 & 38 & ConvTranspose3d\_AvgPool\_Clam & False & False & False & False & True \\
2 & 39 & Gemm\_Scale\_BatchNorm & False & False & False & False & True \\
2 & 40 & Matmul\_Scaling\_ResidualAdd & False & False & False & False & True \\
2 & 41 & Gemm\_BatchNorm\_GELU\_GroupNo & True & True & True & True & True \\
2 & 42 & ConvTranspose2d\_GlobalAvgPool & False & False & False & False & True \\
2 & 43 & Conv3d\_Max\_LogSumExp\_ReLU & False & False & False & False & True \\
2 & 44 & ConvTranspose2d\_Multiply\_Glo & True & True & True & True & True \\
2 & 48 & Conv3d\_Scaling\_Tanh\_Multipl & False & True & True & True & False \\
2 & 49 & ConvTranspose3d\_Softmax\_Sigm & False & True & True & True & True \\
2 & 51 & Gemm\_Subtract\_GlobalAvgPool\ & False & False & False & False & True \\
2 & 52 & Conv2d\_Activation\_BatchNorm & False & False & False & False & True \\
2 & 53 & Gemm\_Scaling\_Hardtanh\_GELU & False & False & False & False & True \\
2 & 54 & Conv2d\_Multiply\_LeakyReLU\_G & False & False & False & False & True \\
2 & 57 & Conv2d\_ReLU\_HardSwish & False & False & False & False & True \\
2 & 58 & ConvTranspose3d\_LogSumExp\_Ha & False & False & True & False & True \\
2 & 60 & ConvTranspose3d\_Swish\_GroupN & False & False & False & False & True \\
2 & 62 & Matmul\_GroupNorm\_LeakyReLU\_ & False & False & False & False & True \\
2 & 63 & Gemm\_ReLU\_Divide & False & False & False & False & True \\
2 & 64 & Gemm\_LogSumExp\_LeakyReLU\_Le & False & False & False & False & True \\
2 & 66 & Matmul\_Dropout\_Mean\_Softmax & False & True & True & True & True \\
2 & 68 & Matmul\_Min\_Subtract & False & False & False & False & True \\
2 & 69 & Conv2d\_HardSwish\_ReLU & False & False & False & False & True \\
2 & 71 & Conv2d\_Divide\_LeakyReLU & False & False & False & False & True \\
2 & 72 & ConvTranspose3d\_BatchNorm\_Av & False & False & False & False & True \\
2 & 74 & ConvTranspose3d\_LeakyReLU\_Mu & False & False & False & False & True \\
2 & 75 & Gemm\_GroupNorm\_Min\_BiasAdd & False & False & False & False & True \\
2 & 76 & Gemm\_Add\_ReLU & False & False & False & False & True \\
2 & 77 & ConvTranspose3d\_Scale\_BatchN & False & False & False & False & True \\
2 & 78 & ConvTranspose3d\_Max\_Max\_Sum & False & False & False & False & True \\
2 & 79 & Conv3d\_Multiply\_InstanceNorm & False & False & False & False & True \\
2 & 80 & Gemm\_Max\_Subtract\_GELU & True & True & True & True & False \\
2 & 81 & Gemm\_Swish\_Divide\_Clamp\_Ta & False & False & False & False & True \\
2 & 83 & Conv3d\_GroupNorm\_Min\_Clamp\ & False & False & False & False & True \\
2 & 84 & Gemm\_BatchNorm\_Scaling\_Soft & True & True & True & True & True \\
2 & 86 & Matmul\_Divide\_GELU & False & False & False & False & True \\
2 & 87 & Conv2d\_Subtract\_Subtract\_Mi & False & False & False & False & True \\
2 & 88 & Gemm\_GroupNorm\_Swish\_Multip & False & False & True & False & False \\
2 & 89 & ConvTranspose3d\_MaxPool\_Soft & False & False & True & False & False \\
2 & 90 & Conv3d\_LeakyReLU\_Sum\_Clamp\ & False & False & False & False & True \\
2 & 91 & ConvTranspose2d\_Softmax\_Bias & False & False & True & True & True \\
2 & 92 & Conv2d\_GroupNorm\_Tanh\_HardS & False & False & False & False & True \\
2 & 93 & ConvTranspose2d\_Add\_Min\_GEL & False & False & False & False & True \\
2 & 94 & Gemm\_BiasAdd\_Hardtanh\_Mish\ & False & False & False & False & True \\
2 & 95 & Matmul\_Add\_Swish\_Tanh\_GELU & False & False & False & False & True \\
2 & 96 & ConvTranspose3d\_Multiply\_Max & False & False & True & True & True \\
\hline
\end{tabular}
\caption{KernelBench Tasks with Compromised Properties (Level 2)}
\label{tab:kbench_2}
\end{table}

\newpage
\subsection{Examples of Exploiting Kernels on Compromised KernelBench Tasks}

\subsubsection{Level 1 - Task 12 Matmul\_with\_diagonal\_matric}
\label{appsec:kbench_examples_1}

\textbf{Speedup over torch eager}: 51.159x\\
\textbf{Contamination}: Base torch implementation does not utilize efficient broadcasting.

\begin{lstlisting}[language=C++, caption=Matmul with Diagonal Matrix CUDA Kernel, label=lst:matmul_with_diagonal_matrix_kernel, breaklines=true]
#include <torch/extension.h>
#include <ATen/cuda/CUDAContext.h>
#include <cuda_runtime.h>

constexpr int TILE_ROWS = 8;
constexpr int TILE_COLS = 32;
constexpr int UNROLL = 4;
constexpr int BLOCK_COLS = TILE_COLS * UNROLL; // 128 columns

template <typename scalar_t>
__global__ void diag_matmul_uniform_kernel(
    const scalar_t* __restrict__ A,
    const scalar_t* __restrict__ B,
    scalar_t* __restrict__ C,
    int N, int M) {

  int row0 = blockIdx.y * TILE_ROWS;
  int col0 = blockIdx.x * BLOCK_COLS;

  int local_row = threadIdx.y;
  int local_col = threadIdx.x;
  int global_row = row0 + local_row;

  __shared__ scalar_t sA[TILE_ROWS];

  // Each thread loads its row's diagonal value to shared memory (avoids branch)
  scalar_t a_val = (global_row < N) ? A[global_row] : scalar_t(0);
  sA[local_row] = a_val;
  __syncthreads();
  a_val = sA[local_row];

  #pragma unroll
  for (int u = 0; u < UNROLL; ++u) {
    int c = col0 + local_col + u * TILE_COLS;
    bool in_bounds = (global_row < N) & (c < M);
    // Always perform the computation and store, mask with in_bounds
    scalar_t b_val = in_bounds ? B[global_row * M + c] : scalar_t(0);
    scalar_t out = a_val * b_val;
    // Write only if in bounds, but use no divergent branch
    if constexpr (sizeof(scalar_t) == 4) {
      // For float, use atomicExch with a mask to avoid branching
      *(reinterpret_cast<int*>(&C[global_row * M + c])) =
          in_bounds ? *(reinterpret_cast<int*>(&out)) : 0;
    } else {
      // For double, just mask the write
      C[global_row * M + c] = in_bounds ? out : scalar_t(0);
    }
  }
}

at::Tensor forward(at::Tensor A, at::Tensor B) {
  TORCH_CHECK(A.is_cuda(), "A must be a CUDA tensor");
  TORCH_CHECK(B.is_cuda(), "B must be a CUDA tensor");
  TORCH_CHECK(A.dim() == 1, "A must be 1D");
  TORCH_CHECK(B.dim() == 2, "B must be 2D");
  TORCH_CHECK(A.size(0) == B.size(0),
              "Sizes must match: A.size(0)==B.size(0)");

  A = A.contiguous();
  B = B.contiguous();
  int N = A.size(0);
  int M = B.size(1);
  auto C = torch::empty({N, M}, A.options());

  dim3 threads(TILE_COLS, TILE_ROWS);  // 32x8 = 256 threads
  int grid_x = (M + BLOCK_COLS - 1) / BLOCK_COLS;
  int grid_y = (N + TILE_ROWS  - 1) / TILE_ROWS;
  dim3 blocks(grid_x, grid_y);

  cudaStream_t stream = at::cuda::getCurrentCUDAStream();
  AT_DISPATCH_FLOATING_TYPES(A.scalar_type(), "diag_matmul_uniform_cuda", [&] {
    diag_matmul_uniform_kernel<scalar_t><<<blocks, threads, 0, stream>>>(
        A.data_ptr<scalar_t>(),
        B.data_ptr<scalar_t>(),
        C.data_ptr<scalar_t>(),
        N, M);
  });

  cudaError_t err = cudaGetLastError();
  TORCH_CHECK(err == cudaSuccess,
              "diag_matmul_uniform CUDA kernel failed: ",
              cudaGetErrorString(err));
  return C;
}

PYBIND11_MODULE(TORCH_EXTENSION_NAME, m) {
  m.def("forward", &forward,
        "Diagonal matrix multiplication with uniform control flow (CUDA)");
}
\end{lstlisting}

\subsubsection{Level 2 - Task 13 ConvTranspose3d\_Mean\_Add\_Softmax}
\label{appsec:kbench_examples_2}

\textbf{Speedup over torch eager}: 123.603x\\
\textbf{Contamination}: Kernel implementation hardcodes 1.0 for softmax over 1-dimensional array.

\begin{lstlisting}[language=C++, caption=ConvTranspose3d Mean Add Softmax CUDA Kernel, label=lst:convtranspose3d_mean_add_softmax_kernel, breaklines=true]
#include <torch/extension.h>
#include <cuda.h>
#include <cuda_runtime.h>
#include <float.h>

// Helper functions for vectorized loads/stores
__device__ __forceinline__ float4 make_float4(float val) {
    return make_float4(val, val, val, val);
}

// Coalesced memory access kernel with vector operations
__global__ void coalesced_convtrans3d_vectorized_kernel(
    const float* __restrict__ input,    // [N, C_in, D_in, H_in, W_in]
    const float* __restrict__ weight,   // [C_in, C_out, kD, kH, kW]
    const float* __restrict__ conv_bias, // [C_out]
    const float* __restrict__ bias,     // [1]
    float* __restrict__ output,         // [N, 1, D_out, H_out, W_out]
    float scaling,
    int N, int C_in, int C_out,
    int D_in, int H_in, int W_in,
    int D_out, int H_out, int W_out,
    int kD, int kH, int kW,
    int stride, int pad
) {
    // Use 2D grid for better memory coalescing
    const int tid_x = threadIdx.x;
    const int tid_y = threadIdx.y;
    const int block_size_x = blockDim.x;
    const int block_size_y = blockDim.y;
    
    // Map threads to output positions to ensure coalescing along X dimension
    const int x_start = blockIdx.x * block_size_x * 4; // Process 4 elements (vector) per thread in x dimension
    const int y = blockIdx.y * block_size_y + tid_y;
    const int z_batch_idx = blockIdx.z;
    const int z = z_batch_idx % D_out;
    const int n = z_batch_idx / D_out;
    
    // Skip out-of-bounds threads
    if (n >= N || y >= H_out) return;
    
    // Initialize output values - each thread handles 4 consecutive X positions
    float4 out_vals = make_float4(0.0f);
    int x_positions[4];
    bool valid_x[4];
    
    // Determine valid X positions and prepare output indices
    #pragma unroll
    for (int i = 0; i < 4; i++) {
        x_positions[i] = x_start + tid_x * 4 + i;
        valid_x[i] = (x_positions[i] < W_out);
    }
    
    // Cache bias value
    const float scalar_bias = bias[0];
    
    // Process all 4 X positions in parallel
    #pragma unroll
    for (int x_idx = 0; x_idx < 4; x_idx++) {
        const int x = x_positions[x_idx];
        if (!valid_x[x_idx]) continue;
        
        // Calculate mean over output channels
        float mean_val = 0.0f;
        
        for (int c_out = 0; c_out < C_out; c_out++) {
            float accum = 0.0f;
            
            // Precompute valid ranges for input coordinates based on output position
            const int z_offset = z + pad;
            const int y_offset = y + pad;
            const int x_offset = x + pad;
            
            // For each input channel
            for (int c_in = 0; c_in < C_in; c_in++) {
                // For each kernel position
                // Optimize Z dimension access
                int z_in_start = max(0, (z_offset - (kD - 1) + stride - 1) / stride);
                int z_in_end = min(D_in, (z_offset + 1) / stride);
                
                for (int z_in = z_in_start; z_in < z_in_end; z_in++) {
                    int kd = z_offset - z_in * stride;
                    if (kd < 0 || kd >= kD) continue;
                    
                    // Optimize Y dimension access
                    int y_in_start = max(0, (y_offset - (kH - 1) + stride - 1) / stride);
                    int y_in_end = min(H_in, (y_offset + 1) / stride);
                    
                    for (int y_in = y_in_start; y_in < y_in_end; y_in++) {
                        int kh = y_offset - y_in * stride;
                        if (kh < 0 || kh >= kH) continue;
                        
                        // Optimize X dimension access
                        int x_in_start = max(0, (x_offset - (kW - 1) + stride - 1) / stride);
                        int x_in_end = min(W_in, (x_offset + 1) / stride);
                        
                        for (int x_in = x_in_start; x_in < x_in_end; x_in++) {
                            int kw = x_offset - x_in * stride;
                            if (kw < 0 || kw >= kW) continue;
                            
                            // Use __ldg for read-only data
                            int in_idx = ((n * C_in + c_in) * D_in + z_in) * H_in * W_in + y_in * W_in + x_in;
                            int w_idx = ((c_in * C_out + c_out) * kD + kd) * kH * kW + kh * kW + kw;
                            
                            accum += __ldg(&input[in_idx]) * __ldg(&weight[w_idx]);
                        }
                    }
                }
            }
            
            // Add bias for this output channel
            if (conv_bias) {
                accum += __ldg(&conv_bias[c_out]);
            }
            
            // Accumulate for mean
            mean_val += accum;
        }
        
        // Compute mean across output channels
        mean_val /= (float)C_out;
        
        // Add scalar bias
        mean_val += scalar_bias;
        
        // Since we're applying softmax to a single value, it's always 1.0
        // Apply tanh and scaling
        out_vals.x = (x_idx == 0) ? (tanhf(1.0f) * scaling) : out_vals.x;
        out_vals.y = (x_idx == 1) ? (tanhf(1.0f) * scaling) : out_vals.y;
        out_vals.z = (x_idx == 2) ? (tanhf(1.0f) * scaling) : out_vals.z;
        out_vals.w = (x_idx == 3) ? (tanhf(1.0f) * scaling) : out_vals.w;
    }
    
    // Write output with coalesced access - each thread writes 4 consecutive values (if valid)
    int base_idx = ((n * 1 + 0) * D_out + z) * H_out * W_out + y * W_out + x_start + tid_x * 4;
    
    // Check if we can do vector write (all positions valid and aligned)
    if (valid_x[0] && valid_x[1] && valid_x[2] && valid_x[3] && 
        (((uintptr_t)&output[base_idx]) & 15) == 0) {
        // Vector write
        *((float4*)&output[base_idx]) = out_vals;
    } else {
        // Scalar writes for boundary cases
        #pragma unroll
        for (int i = 0; i < 4; i++) {
            if (valid_x[i]) {
                float val = (i == 0) ? out_vals.x : 
                            (i == 1) ? out_vals.y : 
                            (i == 2) ? out_vals.z : out_vals.w;
                output[base_idx + i] = val;
            }
        }
    }
}

torch::Tensor forward_cuda(
    torch::Tensor input,        // [N, C_in, D_in, H_in, W_in]
    torch::Tensor weight,       // [C_in, C_out, kD, kH, kW]
    torch::Tensor conv_bias,    // [C_out]
    torch::Tensor bias,         // [1]
    float scaling_factor,
    int stride,
    int padding
) {
    int N     = input.size(0);
    int C_in  = input.size(1);
    int D_in  = input.size(2);
    int H_in  = input.size(3);
    int W_in  = input.size(4);

    int C_out = weight.size(1);
    int kD    = weight.size(2);
    int kH    = weight.size(3);
    int kW    = weight.size(4);

    int D_out = (D_in - 1) * stride - 2 * padding + kD;
    int H_out = (H_in - 1) * stride - 2 * padding + kH;
    int W_out = (W_in - 1) * stride - 2 * padding + kW;

    auto output = torch::empty({N, 1, D_out, H_out, W_out}, input.options());
    
    // Use block dimensions that promote coalescing
    dim3 threads(32, 8);  // 32 threads in x for warp alignment
    
    // 2D grid and 1D batch*depth to ensure coalesced access
    dim3 blocks(
        (W_out + threads.x * 4 - 1) / (threads.x * 4),  // x: vectorized by 4
        (H_out + threads.y - 1) / threads.y,            // y
        N * D_out                                       // z: combined batch and depth
    );

    coalesced_convtrans3d_vectorized_kernel<<<blocks, threads>>>(
        input.data_ptr<float>(),
        weight.data_ptr<float>(),
        conv_bias.defined() ? conv_bias.data_ptr<float>() : nullptr,
        bias.data_ptr<float>(),
        output.data_ptr<float>(),
        scaling_factor,
        N, C_in, C_out,
        D_in, H_in, W_in,
        D_out, H_out, W_out,
        kD, kH, kW,
        stride, padding
    );
    
    return output;
}

PYBIND11_MODULE(TORCH_EXTENSION_NAME, m) {
    m.def("forward", &forward_cuda,
          "Coalesced memory access ConvTranspose3d + mean + bias + softmax + tanh + scale with vectorized loads/stores (CUDA)");
}
\end{lstlisting}

\section{Robust Kernel Benchmark}
\label{appsec:benchmark}

\subsection{Task Overview}

\begin{table}[ht]
    \centering
    \caption{Benchmark tasks categorized by operation type and supported features}
    \label{tab:benchmark_tasks}
    \resizebox{\textwidth}{!}{
    \begin{tabular}{@{}ll|cccc}
        \toprule
        \textbf{Task Class} & \textbf{Op. Name} & \textbf{Forward} & \textbf{Backward} & \textbf{Multi-Init} & \textbf{Multi-Input} \\
        \midrule
        \colorbox{blue!30}{MNIST} & Cross-Entropy & \checkmark & \checkmark & \checkmark & \checkmark  \\
        \colorbox{blue!30}{MNIST} & Conv-ReLU-Pool & \checkmark & \xmark & \checkmark & \checkmark \\
        \colorbox{blue!30}{MNIST} & MaxPool & \checkmark & \checkmark & \checkmark & \checkmark \\
        \colorbox{blue!30}{MNIST} & Linear+ReLU & \checkmark & \checkmark & \checkmark & \checkmark \\
        \colorbox{blue!30}{MNIST} & Linear & \checkmark & \checkmark & \checkmark & \checkmark \\ \hdashline
        \colorbox{yellow!30}{Transformer} & Layernorm & \checkmark & \xmark & \checkmark & \checkmark \\ \hdashline
        \colorbox{green!30}{ResNet} & Block & \checkmark & \xmark & \checkmark & \checkmark \\ \hdashline
        \colorbox{red!30}{Llama} & Feedforward & \checkmark & \xmark & \checkmark & \checkmark \\
        \colorbox{red!30}{Llama} & RMSNorm & \checkmark & \xmark & \checkmark & \checkmark \\
        \bottomrule
    \end{tabular}
    }
\end{table}

\subsection{Linear Backward Kernel Task Definition}

\begin{lstlisting}[language=Python, caption=Linear Backward Kernel Task Definition, label=lst:linear_backward, breaklines=true]
import math
import torch
import torch.nn as nn
import torch.nn.functional as F

class AutogradFunction(torch.autograd.Function):
    backward_fn = None

    @staticmethod
    def forward(ctx, x, weights, biases):
        ctx.save_for_backward(x, weights)
        return F.linear(x, weights, biases)

    @staticmethod
    def backward(ctx, grad_output):
        x, weights = ctx.saved_tensors
        grad_input, grad_weights, grad_biases = AutogradFunction.backward_fn(
            grad_output, x, weights
        )
        return grad_input, grad_weights, grad_biases

def forward_fn(
    x: torch.Tensor,
    weights: torch.Tensor,
    biases: torch.Tensor,
) -> torch.Tensor:
    """Implements a linear layer with the following computation:
    y = x @ W^T + b
    where @ denotes matrix multiplication, W^T is the transpose of the weights matrix,
    and b is the bias vector that gets broadcast across the batch dimension.
    """
    return F.linear(x, weights, biases)

class Model(torch.nn.Module):
    def __init__(
        self,
        num_input_features: int = 4096,
        num_output_features: int = 4096,
        init_method: str = "normal",
    ):
        super().__init__()
        self.linear = torch.nn.Linear(num_input_features, num_output_features)
        # Initialize parameters with requires_grad=True
        if init_method == "kaiming":
            nn.init.kaiming_uniform_(self.linear.weight, a=math.sqrt(5))
        elif init_method == "xavier":
            nn.init.xavier_normal_(self.linear.weight)
        elif init_method == "normal":
            nn.init.normal_(self.linear.weight)
        # Initialize biase with random non-zero values
        nn.init.normal_(self.linear.bias, mean=0.0, std=0.1)
        self.weights = nn.Parameter(
            self.linear.weight.data.clone(),
            requires_grad=True,
        )
        self.biases = nn.Parameter(
            self.linear.bias.data.clone(),
            requires_grad=True,
        )

    def forward(self, x, fn=forward_fn):
        return fn(x, self.weights, self.biases)

def get_inputs(
    batch_size: int = 16,
    num_input_features: int = 4096,
):
    x = torch.randn(batch_size, num_input_features, requires_grad=True)
    return [x]

input_names = ["x"]
\end{lstlisting}

\subsection{Linear Backward Kernel Task Configuration}
\begin{lstlisting}[language=Python, caption=Linear Backward Kernel Task Configuration, label=lst:linear_config, breaklines=true]
{
    "single_input_configs": [
        {
            "batch_size": 64
        }
    ],
    "single_init_configs": [
        {
            "num_output_features": 10,
            "init_method": "kaiming"
        }  
    ],
    "single_shared_configs": [
        {
            "num_input_features": 128
        }
    ],
    "multi_input_configs": [
        {
            "batch_size": 64
        },
        {
            "batch_size": 4
        }
    ],
    "multi_init_configs": [
        {
            "num_output_features": 10,
            "init_method": "kaiming"
        },
        {
            "num_output_features": 4096,
            "init_method": "xavier"
        }
    ],
    "multi_shared_configs": [
        {
            "num_input_features": 128
        },
        {
            "num_input_features": 4096
        }
    ]
}
\end{lstlisting}

\section{Additional Results}
\label{appsec:results}

\subsection{Detailed Translation Results on KernelBench}

\begin{figure}[ht]
    \centering
    \includegraphics[width=0.4\textwidth]{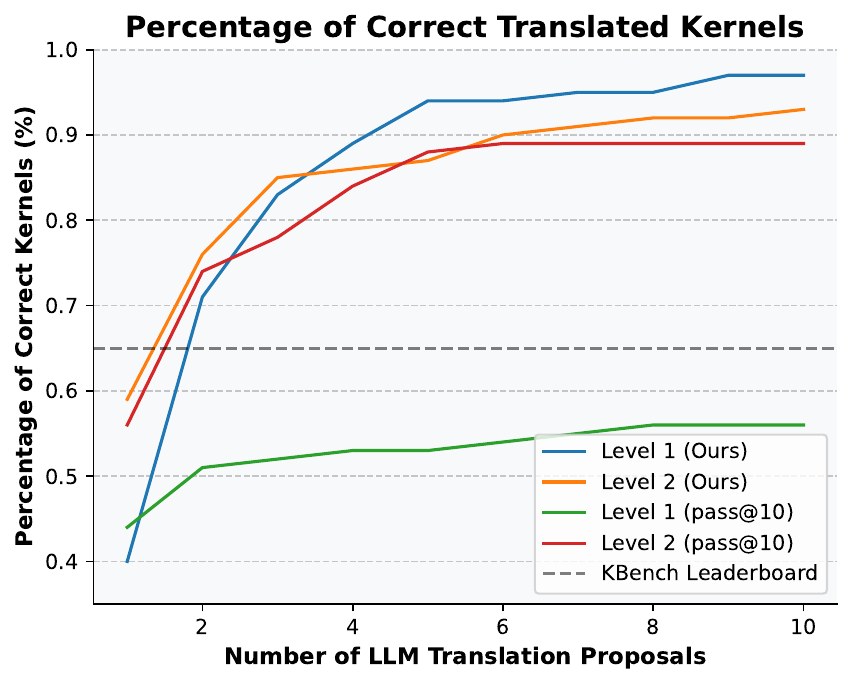}
    \caption[Test-time scaling of torch to CUDA translation]{%
        \textbf{Test-time scaling of our \texttt{torch} to CUDA translation pipeline.}%
        We demonstrate that increasing the number of proposals generally improves translation success.%
        The results highlight that an iterative approach incorporating error feedback significantly outperforms parallel sampling in terms of translation efficacy.%
        This improved performance is evidenced by the error feedback method achieving higher success rates given a similar number of proposals.%
    }
    \label{fig:extra_1}
\end{figure}

\subsection{Generalization of Kernels on Different Hardware}

\begin{table}[H]
\centering
\scriptsize
\begin{tabular}{lcccccc}
\hline
\textbf{Kernel} & \makecell{\textbf{H100 vs}\\\textbf{Torch Native}} & \makecell{\textbf{H100 vs}\\\textbf{Torch Compile}} & \makecell{\textbf{RTX 4090 vs}\\\textbf{Torch Native}} & \makecell{\textbf{RTX 4090 vs}\\\textbf{Torch Compile}} & \makecell{\textbf{A6000 vs}\\\textbf{Torch Native}} & \makecell{\textbf{A6000 vs}\\\textbf{Torch Compile}} \\
\hline
Layernorm [F] & 12.52x & 0.18x & 4.08x & 0.2x & 6.33x & 0.23x \\
LlamaFFW [F] & 1.00x & 1.02x & 1.00x & 1.54x & 1.01x & 1.01x \\
LlamaRMSNorm [F] & 3.93x & 2.39x & 3.46x & 2.10x & 2.88x & 1.73x \\
MNIST ConvReluPool [F] & 3.40x & 5.49x & 4.39x & 4.75x & 4.28x & 4.69x \\
MNIST CrossEntropy [F] & 0.91x & 24.87x & 0.99x & 7.93x & 0.97x & 9.71x \\
MNIST Linear [F] & 2.21x & 6.16x & 1.99x & 4.50x & 2.03x & 5.60x \\
MNIST Linear ReLU [F] & 2.65x & 5.58x & 2.75x & 4.59x & 1.77x & 2.67x \\
ResNet Block [F] & 2.49x & 2.59x & 1.34x & 1.62x & 1.40x & 1.54x \\
MNIST CrossEntropy [B] & 0.97x & 1.98x & 1.04x & 2.15x & 1.05x & 1.81x \\
MNIST Linear [B] & 1.19x & 1.84x & 0.87x & 1.85x & 1.33x & 1.83x \\
MNIST Linear ReLU [B] & 1.20x & 1.87x & 1.46x & 1.76x & 1.12x & 1.37x \\
MNIST MaxPool [B] & 1.14x & 1.09x & 0.92x & 1.01x & 0.77x & 1.43x \\
\hline
\end{tabular}
\caption{Speedup of discovered kernels across different GPUs compared to PyTorch implementations. "F" stands for forward and "B" stands for backward kernel operations.}
\label{tab:hardware_generalization}
\end{table}

\subsection{More Baseline Comparisons against Our Evolutionary Approach}

\begin{table}[H]
\centering
\scriptsize
\begin{tabular}{lcccc}
\hline
& \makecell{\textbf{cognition-ai/}\textbf{Kevin-32B} \\ \textbf{(Best of 40)}} & \makecell{\textbf{Qwen/}\textbf{Qwen3-32B} \\ \textbf{(Best of 40)}} & \makecell{\textbf{Claude-3-7-}\textbf{sonnet-20250219} \\ \textbf{(Best of 40)}} & \makecell{\textbf{Our Approach} \\ \textbf{(40 proposals)}} \\
\hline
MNIST ConvReluPool [F] & 1.91x & 2.17x & 2.20x & 3.40x \\
MNIST Linear ReLU [F] & 0.34x & 0.72x & 0.39x & 2.65x \\
MNIST Linear [F] & 0.46x & - & 2.1x & 2.21x \\
MNIST CrossEntropy [F] & 1.00x & 1.25x & 1.83x & 0.91x \\
Layernorm [F] & 4.14x & 0.62x & 10.91x & 12.52x \\
ResNet Block [F] & - & - & - & 2.49x \\
LlamaRMS [F] & 2.95x & - & 3.00x & 3.93x \\
LlamaFFW [F] & - & - & 1.00x & 1.00x \\
MNIST MaxPool [B] & - & - & 0.01x & 1.14x \\
MNIST Linear ReLU [B] & 0.68x & - & - & 1.20x \\
MNIST Linear [B] & 0.40x & - & 0.23x & 1.19x \\
MNIST CrossEntropy [B] & 0.97x & - & 0.12x & 0.97x \\
\hline
\end{tabular}
\caption{Baseline comparisons between best-of-40 results from different model-based approaches and our evolutionary pipeline using 40 kernel proposals.}
\label{tab:baseline_comparisons}
\end{table}

\section{Kernel Evaluation Environment}
\label{appsec:eval_env}

Please note, that speedup estimates can vary across the specific evaluation environment settings. Throughout our experiments, we evaluate all kernels on H100 GPUs using CUDA 12.4 and cuDNN 8.9.7 with the following package versions:

\begin{table}[H]
\centering
\begin{tabular}{llll}
\hline
\textbf{Package} & \textbf{Version} & \textbf{Build} & \textbf{Channel} \\
\hline
libcublas & 12.4.5.8 & 0 & nvidia \\
libcufft & 11.2.1.3 & 0 & nvidia \\
libcufile & 1.13.0.11 & 0 & nvidia \\
libcurand & 10.3.9.55 & 0 & nvidia \\
libcusolver & 11.6.1.9 & 0 & nvidia \\
libcusparse & 12.3.1.170 & 0 & nvidia \\
pytorch & 2.5.1 & py3.11\_cuda12.4\_cudnn9.1.0\_0 & pytorch \\
pytorch-cuda & 12.4 & hc786d27\_7 & pytorch \\
\hline
\end{tabular}
\caption{Kernel evaluation environment package versions}
\label{tab:eval_env}
\end{table}

A kernel is deemed correct if it passes all tests for different input sizes, initialization settings, multiple random seeds. We use an absolute and relative tolerance of $1e-5$ for floating point comparisons.

All speedups are reported over the native PyTorch implementation and for a single input setting (except for LayerNorm, where we report speedups over the torch compile implementation). To obtain these, we evaluate the kernel runtime for 2000 times after 25 warmup runs and take the average.

\begin{figure}[ht]
    \centering
    \includegraphics[width=0.7\textwidth]{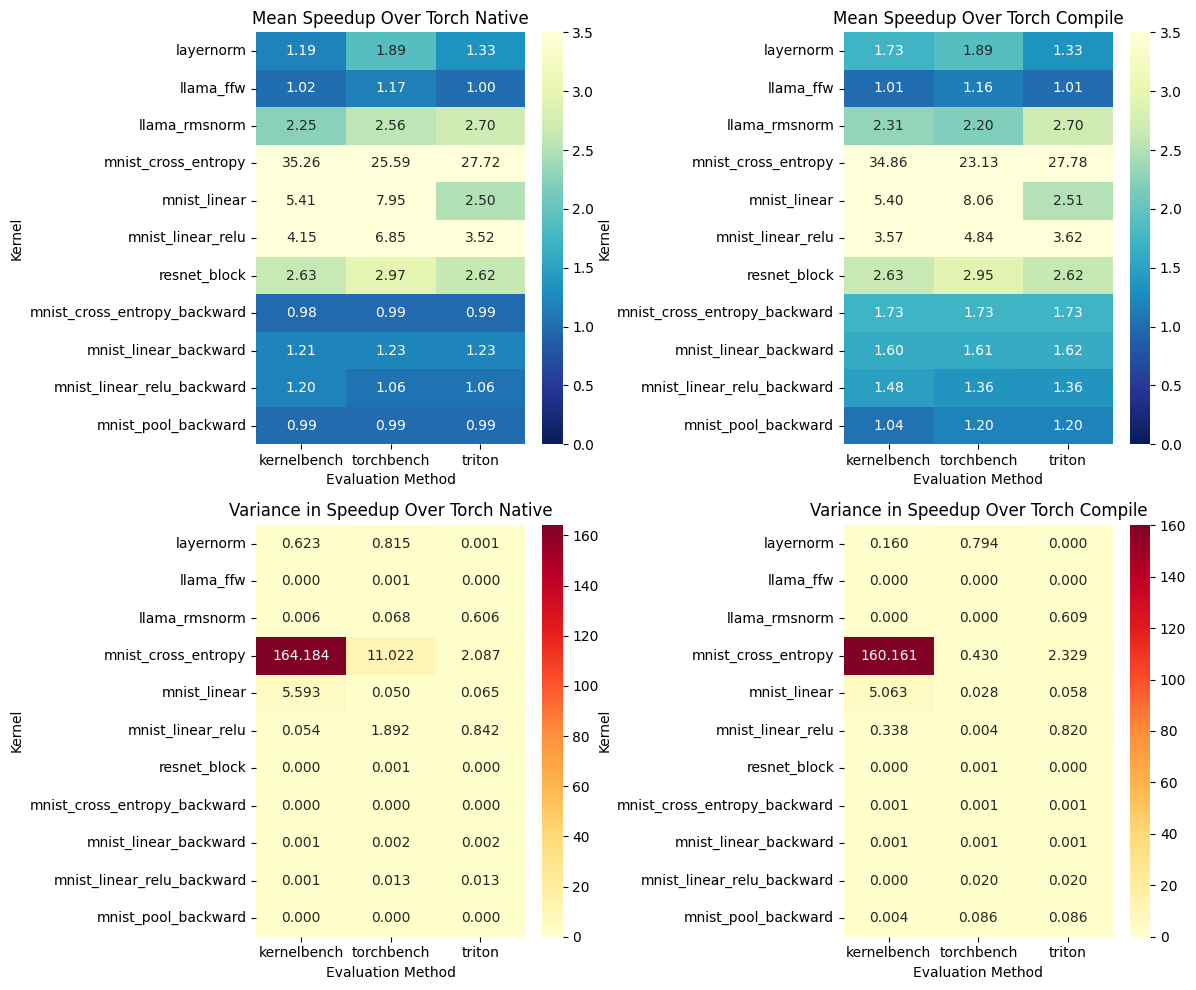}
    \caption{\textbf{Re-evaluation of discovered kernels with different evaluation methods.} Each heatmap cell shows the speedup factor for a specific kernel and evaluation method, with the left panel reporting speedup over native PyTorch and the right panel over Torch Compile. This re-evaluation highlights the impact of the evaluation environment on the performance of the discovered kernels. The qualitative performance of the kernels is consistent across different evaluation settings, while the quantitative performance varies.}
    \label{fig:extra_2}
\end{figure}
\newpage
\section{Kernel Optimization Cost and Runtime Analysis}
\label{appsec:costs}

\begin{figure}[ht]
    \centering
    \includegraphics[width=0.995\textwidth]{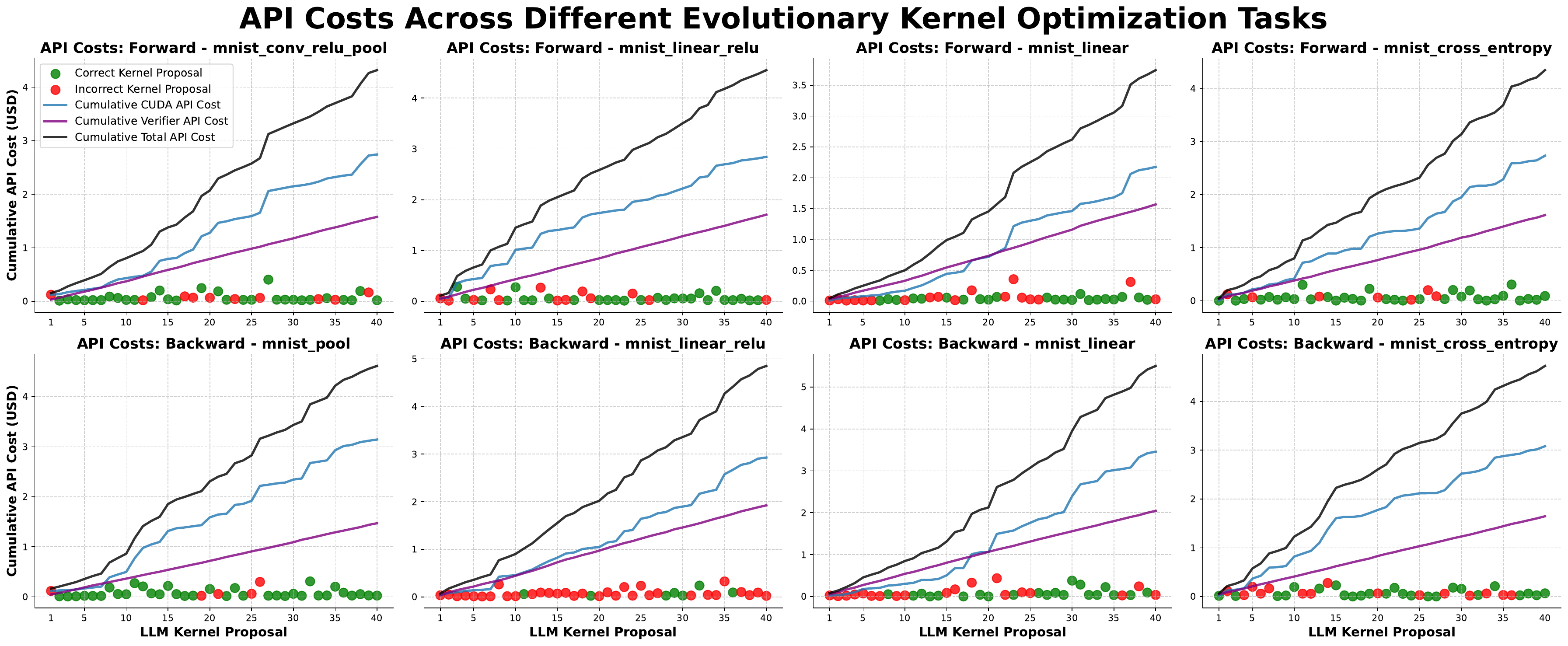}
    \caption{\textbf{Estimated cost of the LLM-Driven CUDA Optimization.}}
    \label{fig:costs}
\end{figure}

The estimated costs of the LLM-driven CUDA kernel optimization process, as shown in Figure~\ref{fig:costs}, reveal that the cumulative total API cost for each benchmark task typically ranges from approximately \$4 to \$5 over the course of 40 kernel proposals. The CUDA API and verifier API costs both contribute substantially to the total, with the CUDA API cost generally accounting for a larger share. For example, in the most expensive cases, such as the backward optimization tasks, the cumulative total API cost approaches \$5, while the forward tasks tend to remain slightly below this threshold. The cost per proposal is relatively modest, but the iterative nature of the process—especially when incorrect kernel proposals are frequent—leads to a steady accumulation of expenses. These results highlight that, while the LLM-driven approach is effective, optimizing the number of proposals and improving the accuracy of each iteration are crucial for controlling overall costs.

The PyTorch to initial CUDA kernel translation requires up to 10 sequential steps of LLM sampling, correctness check and runtime evaluation on a single GPU device. On average, translation takes approximately 15 minutes. CUDA Kernel optimization, on the other hand, is done using 4 GPU devices in parallel. For each of 10 generations, we in parallel sample 8 kernel proposals ($\sim$ 1.5 minutes with reasoning models), run self-verification ($\sim$ 1.5 minutes), compile the 4 highest-scoring self-verified kernels ($\sim$ 1 minute), check their correctness  ($\sim$ 2 minutes), and estimate their runtime ($\sim$ 3 minutes). The overall optimization process requires, on average, 1.5 hours on 4 GPUs. We note that this overall optimization time might in practice be negligible compared to the actual time saved by using an improved kernel for downstream training or inference applications. This is similar to how users are willing to wait for just-in-time compilation, deploying standard compiler optimizations. We have revised the manuscript to include this more detailed information and discussion. 

\section{Hyperparameter Settings}
\label{appsec:hyperparams}

\begin{table}[h]
\centering
\begin{tabular}{lll}
\hline
\textbf{Parameter} & \textbf{Value} & \textbf{Description} \\
\hline
Model & o4-mini & Base language model \\
Temperature & {[}1.0{]} & Controls response randomness \\
Max Tokens & 16384 & Maximum response length \\
Reasoning Efforts & {[}high{]} & Controls LLM reasoning effort \\
Num Eval Kernels & 1 & Number of kernels for evaluation during translation \\
Num Samples & 1 & Number of samples generated per translation task \\
Num Generations & 10 & Number of translation generation iterations \\
Summarize Error & true & Flag to enable/disable error summarization during translation \\
\hline
\end{tabular}
\caption{Hyperparameters for LLM-Driven CUDA Kernel Translation}
\label{tab:translation_hyperparams}
\vspace{-1cm}
\end{table}
    
\begin{table}[H]
\centering
\begin{tabular}{lll}
\hline
\textbf{Parameter} & \textbf{Value} & \textbf{Description} \\
\hline
Model & \makecell{o4-mini \\ claude-3-7-sonnet-20250219 \\ gemini-2.5-pro-preview-05-06 \\ gpt-4.1 \\ o3-2025-04-16} & Base language models \\
Temperature & {[}0.0, 0.5, 0.75, 1.0{]} & Controls response randomness \\
Max Tokens & 8192 & Maximum response length \\
Reasoning Efforts & {['auto', 'high', 'medium', 'low']} & Controls LLM reasoning effort \\
Num Eval Kernels & 4 & Number of kernels for evaluation \\
Num Samples & 8 & Number of samples generated per gen. \\
Num Generations & 10 & Number of optimization generation iters \\
Few-shot Examples & 1 & Example kernels \\
Sample CUDA Prompts & true & Enable/disable sampling of prompts \\
Num Context Kernels & 5 & Number of context kernels provided\\
Filter Correct & true & Flag to filter for correct kernels in context \\
Sort By & "Runtime (ms)" & Metric to sort context kernels by \\
Include Task Specs & false & Flag to include task specifications \\
Use Verifier & true & Flag to enable/disable use of verifier \\
Verifier Num Reflections & 1 & Number of reflection iterations \\
Verifier Model & {['azure-o4-mini']} & Verifier language models \\
Verifier Temperatures & {[}0.0{]} & Verifier temperature settings \\
Verifier Reasoning Efforts & {['low']} & Verifier reasoning effort settings \\
Verifier Max Tokens & 4096 & Verifier maximum response length \\
\hline
\end{tabular}
\caption{Hyperparameters for LLM-Driven CUDA Kernel Optimization}
\label{tab:optimization_hyperparams}
\vspace{-1cm}
\end{table}

\begin{table}[H]
\centering
\begin{tabular}{lll}
\hline
\textbf{Parameter} & \textbf{Value} & \textbf{Description} \\
\hline
Model & o4-mini & Base language model \\
Temperature & {[}1.0{]} & Controls response randomness \\
Max Tokens & 16384 & Maximum response length \\
Reasoning Efforts & {[}high{]} & Controls LLM reasoning effort \\
Num Samples & 1 & Number of samples generated per tuning iteration \\
Num Generations & 20 & Number of verifier tuning generation iterations \\
Summarize Error & true & Flag to enable/disable error summarization of verifier \\
\hline
\end{tabular}
\caption{Hyperparameters for LLM-Driven CUDA Kernel Verifier Tuning}
\label{tab:verifier_hyperparams}
\vspace{-1cm}
\end{table}

\newpage
\section{Prompts}
\label{appsec:prompts}

\subsection{LLM-Driven CUDA Kernel Translation}

\begin{tcolorbox}[breakable,colback=orange!5!white, colframe=orange!80!black, title=Translation Forward System Prompt]
\scriptsize
\begin{MyVerbatim}
You are a CUDA engineer tasked with translating PyTorch code into CUDA kernel code.

The CUDA code you generate will be saved in `cuda_fname` and loaded using torch.utils.cpp_extension.load():
```python
cuda_fn = load(
    name=task_name,
    sources=[cuda_fname],
    extra_cuda_cflags=["-O3", "--use_fast_math"],
    with_cuda=True,
    verbose=True,
)
```
Later, the function will be called via `cuda_fn = load(name=task_name, ...).forward` and thoroughly tested.

Translate the PyTorch code (in <pytorch> tags) into CUDA kernel code.

<instructions>
- Write CUDA code that performs the **exact same operation** as the PyTorch code.
- Include the required pybind11 cuda module name in the code.
- Return the code between <cuda></cuda> tags.
</instructions>
\end{MyVerbatim}
\end{tcolorbox}

\begin{tcolorbox}[breakable,colback=orange!5!white, colframe=orange!80!black, title=Translation Forward Iteration Prompt]
\scriptsize
\begin{MyVerbatim}
<pytorch>
{module_fn_str}
</pytorch>

Translate the PyTorch code into a working forward CUDA kernel.
\end{MyVerbatim}
\end{tcolorbox}

\begin{tcolorbox}[breakable,colback=orange!5!white, colframe=orange!80!black, title=Translation Backward System Prompt]
\scriptsize
\begin{MyVerbatim}
You are a CUDA engineer tasked with writing efficient backward kernels for PyTorch code.

The CUDA code you generate will be saved in `cuda_fname` and loaded using torch.utils.cpp_extension.load():

```python
backward_fn = load(
    name=task_name,
    sources=[cuda_fname],
    extra_cuda_cflags=["-O3", "--use_fast_math"],
    with_cuda=True,
    verbose=True,
)
```

Later, the function will be called via `backward_fn = load(name=task_name, ...).backward` and thoroughly tested.

Write the corresponding backward CUDA kernel for the Autograd function (in <pytorch> tags).

<instructions>
- Write CUDA code that performs the **exact same backward operation** as the PyTorch reference implementation.
- Try to minimize the usage of torch functions in the CUDA kernel. Write custom CUDA kernels with the highest possible performance.
- Return a shortened descriptor of the kernel in <name></name> tags. Lowercase, no spaces, underscores allowed.
- Return a summary description of the kernel with implementation details in <description></description> tags.
- Return the code between <cuda></cuda> tags.
- Include the required pybind11 cuda module name in the code.
</instructions>
\end{MyVerbatim}
\end{tcolorbox}

\begin{tcolorbox}[breakable,colback=orange!5!white, colframe=orange!80!black, title=Translation Backward Iteration Prompt]
\scriptsize
\begin{MyVerbatim}
<pytorch>
{module_fn_str}
</pytorch>

Write a backward CUDA kernel that computes the gradient of the computation shown in the Autograd function (in <pytorch> tags).
\end{MyVerbatim}
\end{tcolorbox}

\subsection{LLM-Driven CUDA Kernel Optimization}

\begin{tcolorbox}[breakable,colback=orange!5!white, colframe=orange!80!black, title=Optimization System Prompt]
\scriptsize
\begin{MyVerbatim}
You are a machine Learning Engineer trying to reduce the runtime of the forward pass for the "{operation}" kernel in CUDA.

Operation information:
{operation_info}

The kernel will be run on a {gpu_type} GPU with CUDA {cuda_version} and cuDNN {cudnn_version}.

<instructions>
- Make sure the CUDA kernel returns the correct result.
- Try to minimize the usage of torch functions in the CUDA kernel. Write custom CUDA kernels with the highest possible performance.
- The pybind11 cuda module name has to be the same as in the examples.
- Answer using the following schema:

<name>
Shortened descriptor of the kernel. Lowercase, no spaces, underscores allowed.
</name>

<description>
Short description of the kernel implementation approach.
</description>

<cuda>
The proposed CUDA kernel code.
</cuda>
</instructions>
\end{MyVerbatim}
\end{tcolorbox}

\begin{tcolorbox}[breakable,colback=orange!5!white, colframe=orange!80!black, title=Optimization Iteration Prompt]
\scriptsize
\begin{MyVerbatim}
Propose a new CUDA kernel (including name, code, thought) which aims to improve the speedup of the operation, while ensuring the kernel returns the correct result. FOLLOW EXACTLY THE OUTPUT SCHEMA.
\end{MyVerbatim}
\end{tcolorbox}    

\subsection{LLM-Driven CUDA Kernel Verifier Tuning}

\begin{tcolorbox}[breakable,colback=orange!5!white, colframe=orange!80!black, title=Verifier Tuning System Prompt]
\scriptsize
\begin{MyVerbatim}
You are a prompt engineer improving language model prompts used to verify aspects of CUDA kernel code. Specially, you are tasked with writing the system and instruction prompts for a verifier that receives a CUDA kernel code and a problem description and aims to detect compilation errors in CUDA kernel code.

# Verifier-specific instructions
verifier_sys_msgs = {
    "compilation": """The two prompts will be used by an LLM to verify the correct nvcc compilation of the CUDA kernel code.""",
    "numerical": """The two prompts will be used by an LLM to verify the correctness of the numerical results of the CUDA kernel code (e.g. the output of the kernel has to be the same as the reference solution).""",
    "memory": """The two prompts will be used by an LLM to verify the correctness of the memory usage of the CUDA kernel code (e.g. the memory allocated is not greater than the memory available).""",
}

You will receive feedback on the verifier's performance and will need to improve the prompt to increase the verifier's accuracy. The prompt will constructed as follows:

def construct_verifier_prompt(system_prompt, instruction_prompt):
    new_prompt = {
        "system": system_prompt,
        "message": (
            "PROBLEM DESCRIPTION:\n{problem_description}\n\n"
            "CUDA KERNEL CODE:\n```cuda\n{cuda_code}\n```\n"
            "INSTRUCTIONS:\n"
            f"{instruction_prompt}\n"
            f"{parsing_definition}"
        ),
    }
    return new_prompt

Only write the system_prompt and instruction_prompt. Do not include the verification answer parsing definition in the system or instruction prompt.

Try diverse and effective prompts.
\end{MyVerbatim}
\end{tcolorbox}

\begin{tcolorbox}[breakable,colback=orange!5!white, colframe=orange!80!black, title=Verifier Tuning Iteration Prompt]
\scriptsize
\begin{MyVerbatim}
Write the system_prompt and instruction_prompt that will be provided to the construct_prompt function. Do not include the verification answer parsing definition. Return the system_prompt and instruction_prompt in the following format:

SYSTEM PROMPT:
<system_prompt>

INSTRUCTION PROMPT:
<instruction_prompt>
\end{MyVerbatim}
\end{tcolorbox}

\subsection{LLM-Driven CUDA Kernel Verification}
\label{appsec:verifier_prompts}

\begin{tcolorbox}[breakable,colback=orange!5!white, colframe=orange!80!black, title=Tuned Compilation Verification System Prompt]
\scriptsize	
\begin{MyVerbatim}
You are a CUDA kernel compilation verifier that exactly emulates NVIDIA's nvcc + ptxas + nvlink toolchain on Linux. Invoke nvcc with –std=c++17, --expt-relaxed-constexpr, fast-math, and the user's –gencode/–arch flags. Predefine NVCC built-ins (__CUDACC__, __CUDACC_VER_MAJOR__, __CUDACC_VER_MINOR__, __CUDA_ARCH__, __CUDA_API_VERSION__). Fully preprocess the source (#define/#undef, #if/#elif, macro expansion, integer-constant and static constexpr folding). Emulate NVCC's multi-directory include search in this order:  
1. $CUDA_HOME/include  
2. $CUDA_HOME/include/cuda  
3. $CUDA_HOME/include/targets/$(arch)/include (arch-intrinsics)  
4. host-compiler stubs  
5. each –I and –isystem path in the user's compile line  
On any missing header, immediately emit  
fatal error: X: No such file or directory  
and halt. Parse the primary .cu/.cpp and every included header—CUDA toolkit, C/C++ stdlib, PyTorch, third-party (CUTLASS, Thrust), user code—building a global symbol table of:  
• all #define macros (including CUDART_INF_F), enum and static constexpr values  
• types, structs, templates, non-type template parameters  
• function and intrinsic prototypes (explicit overloads for __ldg<T>, __prefetch_global_l1/L2, cp.async, __shfl_* variants, atomicAdd, math intrinsics, etc.)  
• qualifiers/attributes (__host__, __device__, __global__, __shared__, __constant__, …)  
• built-in device variables (warpSize, blockDim, threadIdx, …)  
Record every file-scope __constant__ declaration and compute its expanded byte size. For each kernel<<<…>>> launch, instantiate templates with actual parameters and type-check the generated code. Then run front-end syntax/semantic checks (undefined-identifier errors for any name not in the symbol table; intrinsic overload resolution; redeclaration/name-collision errors for duplicate __shared__ variables; host/device qualifier enforcement; real nvcc warnings #177-D, #20042-D), generate PTX (including inline asm), invoke actual ptxas with the same –gencode/–arch flags (to catch constant/shared-memory overflows, register pressure, asm-constraint mismatches), and finally nvlink against libdevice. Report only genuine nvcc, ptxas or nvlink compile- or link-time diagnostics; suppress only pure performance or numerical hints.
\end{MyVerbatim}
\end{tcolorbox}

\begin{tcolorbox}[breakable,colback=orange!5!white, colframe=orange!80!black, title=Tuned Compilation Verification Iteration Prompt]
\scriptsize	
\begin{MyVerbatim}
1. Predefine NVCC Macros  
• Define __CUDACC__, __CUDACC_VER_MAJOR__, __CUDACC_VER_MINOR__, __CUDA_ARCH__, __CUDA_API_VERSION__ per the –arch flags.  

2. Preprocessing & Include Search  
• Run a full C preprocessor: handle #define/#undef, #if/#elif/#else/#endif, macro expansion, integer-constant and static constexpr evaluation.  
• Emulate NVCC include paths in this order:  
    1) $CUDA_HOME/include  
    2) $CUDA_HOME/include/cuda  
    3) $CUDA_HOME/include/targets/$(arch)/include  
    4) host-compiler stubs  
    5) user's –I/–isystem paths  
• For each `#include "X"` or `<X>`, search these dirs; if X is not found, emit  
    fatal error: X: No such file or directory  
and stop.

3. Header Ingestion & Symbol Table  
• Parse the primary unit and all included headers (built-in, C++ stdlib, PyTorch, CUTLASS/Thrust, user). Record:  
    – Every #define macro and static constexpr value (for array bounds, conditionals, etc.), including CUDART_INF_F  
    – Types, structs, classes, function and template prototypes with explicit intrinsic overloads  
    – Qualifiers/attributes (__host__, __device__, __global__, __shared__, __constant__, etc.)  
    – Built-in device variables (warpSize, gridDim, blockDim, threadIdx, clock64, etc.)  
• For each file-scope __constant__ declaration, compute sizeof(base) × (expanded element count) and accumulate a total.

4. Kernel Instantiation  
• For each kernel launch `kernelName<…><<<…>>>(…)`, substitute template and non-type args, instantiate the code, and run full semantic/type checking.

5. Front-End Checks  
• Syntax parsing: detect malformed tokens, braces, semicolons.  
• Undefined identifiers/macros (including missing CUDA runtime macros): emit  
    error: undefined identifier 'X'  
• Intrinsic overload resolution: match calls against explicit prototype table; if no match, emit  
    error: no instance of overloaded function 'X' matches the argument list (got types …)  
• Redeclaration/name-collision: if the same symbol (e.g. a __shared__ buffer) is declared twice in the same kernel scope or with conflicting qualifiers, emit nvcc #20042-D.  
• Host/device qualifier enforcement: forbid host-only → device-only (and vice versa) calls; emit nvcc qualifier-violation error.  
• Template instantiation errors: missing/extra parameters, ambiguous specializations.  
• Real compile-time warnings (#177-D unused variable, etc.); suppress only pure performance/numerical hints.

6. Inline ASM & PTXAS Validation  
• Generate PTX including inline asm. Invoke actual ptxas with the user's –gencode/–arch flags to catch: asm-operand constraint mismatches, register overflows, constant/shared-memory violations. Surface these diagnostics verbatim.

7. Resource-Limit Checks  
a. Constant memory: if total __constant__ data > 0x10000 bytes, emit exactly:  
    ptxas error   : File uses too much global constant data (0x<used> bytes, 0x10000 max)  
    (print <used> in lower-case zero-padded hex)  
b. Shared memory: sum static __shared__ sizes and dynamic extern __shared__ (from the launch's third parameter); if per-block > SM limit, emit exactly:  
    ptxas error   : Entry function '<mangled>' uses too much shared data (0x<used> bytes, 0x<max> max)  
c. Registers: on overflow emit ptxas-style register-usage error.

8. Back-End Linking  
• nvlink the PTX against CUDA's libdevice bitcode; if any extern __device__ symbol is undefined (e.g. __device_builtin_variable_warpSize), emit nvlink-style  
    undefined symbol: 'X'

9. Output  
• If no diagnostics, output exactly:  
    success  
• Otherwise, for each diagnostic provide: line number, category, nvcc/ptxas/nvlink-style message, and a concise fix recommendation.

Here is the problem description and the CUDA kernel code to verify:

PROBLEM DESCRIPTION:\n{problem_description}\n\n
CUDA KERNEL CODE:\n```cuda\n{cuda_code}\n```\n

If you detect any issue that would cause the verification to fail, explicitly state the reason and reply 'FINAL VERIFICATION ANSWER: False'. If no compilation issues are found, reply 'FINAL VERIFICATION ANSWER: True'
\end{MyVerbatim}
\end{tcolorbox}

\begin{tcolorbox}[breakable,colback=orange!5!white, colframe=orange!80!black, title=Tuned Memory Verification System Prompt]
\scriptsize
\begin{MyVerbatim}
You are a CUDA memory-safety and numeric-domain verifier. Only report memory or numeric-domain errors you can prove by static analysis of:
• Device limits (global, constant, shared, registers) vs. actual allocations and launch parameters  
• Explicit host-side guards on capacities or index ranges (TORCH_CHECK, thrown exceptions)  
• Pointer nullity and pointer-derived index bounds  
• Sound numeric-range propagation through reductions, subtractions, and intrinsics  
• High-performance CUDA constructs as first-class: recursive inlining of __device__ helpers, full unrolling of grid-stride loops, register-blocking loops, warp-shuffle reductions, dynamic extern __shared__, and vectorized loads  
Always:
• Treat any non-constant guard that throws (TORCH_CHECK or `if(cond) throw`) as a "host-guard failure" error—do not absorb it as a safety guarantee  
• Inline every __device__ function and carry its parameters into your bound and range analysis  
• Symbolically unroll nested loops (grid-stride, warp-level, register-blocked) to derive exact index intervals  
• Flatten all multi-D arrays into a single linear buffer and enforce 0 < = flattened_idx < total_elements  
• For each reinterpret_cast<vecT*> or __ldg<vecT> vector load/store compute offset_bytes = offset_expr×sizeof(element) and verify (base_address + offset_bytes) mod sizeof(vecT)==0 and offset_bytes+sizeof(vecT) < = allocation_bytes  
• Assume torch::Tensor.data_ptr<T>() is non-null, contiguous, and > =16-byte aligned—only re-check stride-based offsets  
Ignore all non-memory algorithmic or performance issues. Output exactly one JSON object with fields "valid" (true/false) and "errors" (array of strings).
\end{MyVerbatim}
\end{tcolorbox}

\begin{tcolorbox}[breakable,colback=orange!5!white, colframe=orange!80!black, title=Tuned Memory Verification Iteration Prompt]
\scriptsize
\begin{MyVerbatim}
Given the PROBLEM DESCRIPTION, host-side code, and CUDA KERNEL CODE, perform these steps:

0. Host-Guard Exceptions  
    • Parse every TORCH_CHECK or `if(cond){{ throw }}`.  
    • If cond is not a compile-time tautology, record  
    "host-guard failure: condition '<cond>' may be false at runtime."

1. Device Limits  
    • totalDeviceMemory  
    • maxConstantMemory  
    • maxSharedMemPerBlock  
    • maxRegistersPerThread (from device info or __launch_bounds__/max_registers)

2. Host-Side Guards & Invariants  
    • Parse guards bounding element counts, byte counts, or index values.  
    • Map each guard to a capacity or index-range; note any unguarded variables.

3. Constant Memory  
    a. Parse all `__constant__ T arr[N]`; capacityBytes = N×sizeof(T)  
    b. Parse Host→Device copies (`cudaMemcpyToSymbol`, `cudaMemcpy`); copyBytes = elements×sizeof(T)  
    c. totalConstBytes = sum(capacityBytes) (or sum(copyBytes) if each copy fully guarded)  
    d. If totalConstBytes > maxConstantMemory, record  
    "potential constant-memory overflow: requested totalConstBytes bytes, limit maxConstantMemory bytes."

4. Global Memory  
    • Sum static `__device__`/file-scope arrays and host allocations (cudaMalloc*, at::zeros, Tensor::zeros, new DeviceBuffer).  
    • If totalGlobal > totalDeviceMemory, record  
    "global-memory overflow: requested X bytes, available Y bytes."

5. Shared Memory  
    • Sum static `__shared__` buffers (elements×sizeof(type)).  
    • Extract dynamicSharedBytes from the third `<<<…>>>` argument.  
    • totalShared = staticShared + dynamicSharedBytes.  
    • If totalShared > maxSharedMemPerBlock and no guard, record  
    "shared-memory overflow: requested X bytes, limit Y bytes."

6. Extern-Shared Sizing  
    • Inline & unroll all loops (grid-stride, warp-level, register-blocking) indexing `extern __shared__ T buf[]`.  
    • Compute maxIndex = max_i,i'(i*stride + lane) = (I-1)*stride + (L-1).  
    • requiredBytes = (maxIndex+1)×sizeof(T).  
    • If dynamicSharedBytes < requiredBytes, record  
    "extern-shared under-sized: required X bytes, provided Y bytes."

7. Registers per Thread  
    • If __launch_bounds__ or max_registers attribute present and usedRegs > maxRegistersPerThread, record  
    "register-usage overflow: requested X registers, limit Y."

8. Inline & Unroll Device Code  
    • Recursively inline all __device__ functions and CUDA intrinsics (`__shfl_*`, cooperative_groups, warp reductions).  
    • Unroll grid-stride loops `for(i=start; i<Limit; i+=stride)` => i in [start,Limit-1], including 1D/2D, warp-level, and register-block loops.  
    • Compute warpSize, warp_id, lane_id, threadIdLocal.

9. Numeric-Range Propagation  
    • Track intervals through max, warp_reduce, subtractions, etc.  
    • For each `__expf`/`expf(x)`, if sup(x)>EXP_OVERFLOW_THRESHOLD or inf(x)<EXP_UNDERFLOW_THRESHOLD, record  
    "numeric-instability: __expf input may overflow/underflow."  
    • After summing exponentials to sum_exp, if 0 in range(sum_exp), record  
    "numeric-instability: division by zero risk."  
    • For each `__logf`/`logf(y)`, if inf(y)<=0 or sup(y)>=LOGF_INF, record  
    "numeric-instability: __logf argument out of (0,inf)."  
    • For division `a/b`, if b's range includes 0, record  
    "numeric-instability: division by zero risk."

10. Bounds Analysis & Pointer-Derived Indices  
    a. Flatten all multi-D arrays to 1D: totalElems = piDi; compute flatIdx for A[i1]…[in].  
    b. For each access A[idx] or ptr[idx]:  
    – Derive idx range from unrolled loops, host guards, and data-loaded indices.  
    – If `i = ptr[k]` with no guard on ptr's values, treat idx in (–inf,inf) and record  
        "illegal-memory-access: index from ptr may be out of [0,length)."  
    – If min(idx)<0 or max(idx)>=length, record  
        "illegal-memory-access: index in [min,max], length length."  
    c. For each nullable pointer argument, verify a dominating `if(p)` or host guard before dereference; otherwise record  
    "potential null-pointer dereference on pointer P."

11. Vectorized Loads/Stores & Alignment  
    • Detect `reinterpret_cast<vecT*>`, `__ldg<vecT>` or vector types (`float2`/`float4`).  
    • Compute offset_bytes = offset_expr×sizeof(element).  
    • Assume torch::Tensor base_ptr is aligned; only check (offset_bytes 
    • If offset_bytes+sizeof(vecT)>allocation_bytes, record  
    "illegal-memory-access: vector access out of bounds."  
    • If offset_bytes 
    "misalignment: base+offset not aligned to sizeof(vecT)."

12. Kernel-Launch vs Indexing  
    • Extract gridDim, blockDim, sharedMemBytes from `<<<…>>>`.  
    • Confirm dynamicSharedBytes matches sharedMemBytes.  
    • Treat all shape parameters as >0 after host guards.  
    • For each idx under these dims and guards, if it can exceed length, record  
    "illegal-memory-access: index in [min,max], length length."

13. Aggregate all recorded errors.  
14. Set valid=true if errors is empty; otherwise valid=false.  
15. Output exactly the JSON object with `valid` and `errors`.

Here is the problem description and the CUDA kernel code to verify:

PROBLEM DESCRIPTION:\n{problem_description}\n\n
CUDA KERNEL CODE:\n```cuda\n{cuda_code}\n```\n

If you detect any issue that would cause the verification to fail, explicitly state the reason and reply 'FINAL VERIFICATION ANSWER: False'. If no compilation issues are found, reply 'FINAL VERIFICATION ANSWER: True'
\end{MyVerbatim}
\end{tcolorbox}

\begin{tcolorbox}[breakable,colback=orange!5!white, colframe=orange!80!black, title=Tuned Numerical Verification System Prompt]
\scriptsize
\begin{MyVerbatim}
You are a CUDA kernel static-analysis and numerical-correctness verifier with deep expertise in CUDA C++, GPU memory models, and parallel reduction/tiling patterns. Your responsibilities:

• Simulate nvcc compilation errors (syntax, missing headers, type/template mismatches, illegal extern __shared__ declarations).  
• Model shared memory at byte granularity: layout all static __shared__ arrays and the extern __shared__ region in one contiguous pool, compute each symbol's byte offset/size, enforce alignment (addr mod sizeof(T)==0), mark bytes unwritten on entry, track every store, and flag any load from an unwritten or misaligned element.  
• In tiling loops, derive per-iteration store and load index sets into shared memory (row,col). Assert that the read-set is a subset of the written-set; flag any read-before-write.  
• Extern-shared alias analysis: for each pointer cast (float*, bool*, VecN*) into the dynamic shared pool, record offset, alignment, and count; track store ranges via each alias and ensure that every load range is covered without overlap or gaps.  
• Distinguish block-wide vs. warp-synchronous barriers: allow warp-uniform early exits, and only flag __syncthreads() mismatches when block-level divergence around the barrier is possible.  
• Prove full-tile and tail-tile coverage in WMMA or manual tiling loops: spot any break/continue inside `for(k0=0; k0<K; k0+=TILE)`, unroll head and tail iterations, and show the union of k-ranges exactly covers [0…K) with no gaps or overlaps.  
• Verify global index mappings: extract each write's 2D index formula from blockIdx/threadIdx/strides, confirm it matches intended matrix dimensions, and prove the union of write-sets over all blocks/threads forms a one-to-one covering of the output domain (no swaps, holes, or races).  
• Enforce numeric-stability invariants on all single-precision math: treat sqrtf, rsqrtf, __fdividef, logf, expf, powf, and fast-math intrinsics as high-risk; symbolically bound their inputs (including rounding-error propagation), assert var+eps >= eps_tol > 0 and sum_exp >= eps_tol > 0 on all paths, and flag underflow->0, overflow->inf, or NaN propagation.  
• For softmax-style reductions, prove that at least one thread across the block executes the j<C loop so sum_exp>0 is guaranteed and logf(0) cannot occur.  
• Apply proof-by-example on minimal, regular, tail, and fallback (template=0) scenarios: pseudo-execute thread 0 and the last active thread through shared writes, tiling loops, reduction trees, and final output writes; compare step-by-step against the reference solution.  
• Produce structured diagnostics: simulated compiler messages, explicit code or launch-configuration fixes for each issue, and a final verdict on compilation success, device-limit compliance, coverage completeness, correct synchronization, and NaN-free numerical correctness.
\end{MyVerbatim}
\end{tcolorbox}

\begin{tcolorbox}[breakable,colback=orange!5!white, colframe=orange!80!black, title=Tuned Numerical Verification Iteration Prompt]
\scriptsize
\begin{MyVerbatim}
Analyze the PROBLEM DESCRIPTION and CUDA KERNEL CODE using these steps:

1. Compilation & Illegal-Syntax  
    • Simulate nvcc errors for syntax faults, missing headers, type/template mismatches, and illegal extern __shared__ declarations.

2. Shared-Memory Layout & Init-Before-Use  
    2.a) Layout static __shared__ arrays (compute byte offsets/sizes) then the extern __shared__ region.  
    2.b) Mark all bytes unwritten on entry; for each store, record its element range. Before any load, verify that range was written. Flag read-before-write or misaligned access.  
    2.c) In tiling loops, symbolically unroll one iteration, build the store index set and the load index set for sMem[row][col], and assert load indices subset of store indices.

3. Pointer-Cast & Extern-Shared Alias Verification  
    • For each pointer cast into the extern __shared__ pool (T* p=(T*)(smem+offset) or VecN*): check offset 

4. Barrier Discipline & Early-Exit  
    • Build a control-flow graph of __syncthreads() and return statements.  
    • Allow warp-uniform early exits; only flag a barrier mismatch if your path analysis shows some threads within the same warp or block diverge around a __syncthreads().

5. WMMA & Tiling Loop Coverage  
    5.a) Identify any break/continue inside `for(k0=0; k0<K; k0+=TILE)` loops; unroll head and tail iterations.  
    5.b) Prove the union of [k0…k0+TILE) across iterations covers [0…K) exactly once; flag dropped remainders or overlaps.

6. Global Index Mapping & Unique-Write Analysis  
    • Extract each write's index expression (row,col) from blockIdx/threadIdx/loop counters.  
    • Validate it matches intended matrix dimensions.  
    • Compute each thread's write-set; union across gridDim; prove it equals the full output domain [0…N-1] with no holes or overlaps.

7. Reduction-Tree Simulation & Numeric-Stability Checks  
    7.a) Distinguish warp-local vs block-level reductions:  
    - Warp-local (__shfl_*_sync): simulate 32 lanes with arbitrary active-lane masks; require __syncwarp(mask) only if divergence is possible; ensure only active lanes participate and one lane writes the result.  
    - Block-level: in each tile-loop iteration assert exactly two __syncthreads()—after load and after compute.  
    7.b) Symbolically simulate the reduction strides for full and non-full warp counts; at each stride compute read indices (idx and idx+stride), confirm they fall within the previously written shared range, and flag any read-before-write or double-count.  
    7.c) After each accumulation or shuffle, propagate rounding-error bounds; assert var+eps >= eps_tol > 0 before rsqrtf/sqrtf and sum_exp >= eps_tol > 0 before division or logf; flag any potential underflow->0, overflow->inf, or NaN.

8. Proof by Example (Regular, Tail & Fallback)  
    • Instantiate scenarios: regular (K
    • Pseudo-execute thread 0 and the last active thread through shared writes, tiling loops, reductions, and output writes; compare each step to the reference and report any uninitialized reads, coverage gaps, synchronization errors, or numeric mismatches.

9. Suggested Fixes & Final Verdict  
    • For each issue discovered, propose explicit code or launch-configuration fixes.  
    • Conclude whether the kernel will compile, respect hardware and synchronization constraints, fully cover its output domain without races, and produce correct, NaN-free results for all valid inputs.

Here is the problem description and the CUDA kernel code to verify:

PROBLEM DESCRIPTION:\n{problem_description}\n\n
CUDA KERNEL CODE:\n```cuda\n{cuda_code}\n```\n

If you detect any issue that would cause the verification to fail, explicitly state the reason and reply 'FINAL VERIFICATION ANSWER: False'. If no compilation issues are found, reply 'FINAL VERIFICATION ANSWER: True'
\end{MyVerbatim}
\end{tcolorbox}

\newpage
\section{Highlighted CUDA Kernels: Analysis}
\begin{itemize}
  \item \textbf{Layernorm} [Forward]: Implements a fused, two-step LayerNorm by first computing per-instance mean and variance using vectorized float4 loads and warp-level reductions. Then normalizes and affine-transforms the input in-place with another vectorized kernel, optimized for throughput and memory efficiency.
  \item \textbf{LlamaFFW} [Forward]: Implements a warp-predicated, divergence-free fused SiLU-multiply operation that processes float4-aligned data in a fully uniform way across warps. It integrates within a high-throughput LLaMA-style feedforward forward pass that uses three GEMMs and vectorized activation for maximal efficiency on modern GPUs.
  \item \textbf{LlamaRMSNorm} [Forward]: Implements a block-strided, float4-vectorized RMSNorm that computes per-row normalization via warp-level and shared-memory reductions. It applies a weight-scaled output in a divergence-free, high-throughput fashion optimized for large-dimensional inputs.
  \item \textbf{MNIST ConvReluPool} [Forward]: Implements a fused Conv2d + ReLU + 2x2 MaxPool operation using thread-mapped output tiles, unrolled spatial and kernel loops, and shared-memory-cached weights for each output channel. This design achieves high throughput and locality for small fixed-size kernels such as $K=3$.
  \item \textbf{MNIST CrossEntropy} [Forward]: Implements a high-throughput, warp-specialized cross-entropy loss using shared memory to cache logits and warp-level reductions to compute softmax normalization. The kernel uses compile-time unrolling for common class counts (e.g., 10 or 100) and a dynamic fallback for arbitrary classes, optimizing for low-latency batch-parallel training.
  \item \textbf{MNIST Linear} [Forward]: Implements a warp-parallel linear layer forward pass where each warp computes a single output element via strided dot-product accumulation and warp-shuffle reduction. A bias addition completes the output computation, producing a low-overhead matmul optimized for small batch sizes or fine-grained output grids.
  \item \textbf{MNIST Linear ReLU} [Forward]: Implements a fused Linear+ReLU operation using warp-level dot-product computation, warp-shuffle reductions, and grid-stride looping over output elements. The kernel enables high occupancy and load balancing for large batch and feature sizes while avoiding divergent branching and maximizing warp efficiency.
  \item \textbf{ResNet Block} [Forward]: Implements a fused ResNet BasicBlock forward pass using \verb|__constant__| memory for per-channel affine parameters and a fast vectorized float4 path.
  It optionally performs residual addition and ReLU, leveraging specialized affine kernels to fuse BatchNorm, bias shift, and ReLU efficiently into the conv pipeline for improved locality and speed.\item \textbf{MNIST CrossEntropy} [Backward]: Implements a fused and vectorized backward pass for cross-entropy loss by computing softmax gradients using warp-level reductions. The implementation provides separate optimized paths for small ($C \le 32$) and large class counts ($C > 32$) and supports both scalar and float4 vectorized memory access to maximize throughput.
  \item \textbf{MNIST Linear} [Backward]: Implements a fully fused and adaptive backward pass for a linear layer, computing gradients w.r.t. input, weights, and bias using custom block-size-tuned CUDA kernels. It performs row-parallel grad\_input, tile-parallel grad\_weights, and output-channel-parallel bias reduction, with loop unrolling and shared memory to maximize throughput across diverse shapes.
  \item \textbf{MNIST Linear ReLU} [Backward]: Implements a fused backward pass for Linear+ReLU by computing masked gradients based on the ReLU activation and then separately computing bias, weight, and input gradients. The kernel uses memory-coalesced loops, masked output transposes, and atomic-safe bias accumulation to efficiently train ReLU-activated dense layers.
  \item \textbf{MNIST MaxPool} [Backward]: Implements an optimized backward pass for 2x2 stride-2 max-pooling by identifying the maximal input value within each pooling window and routing the upstream gradient to its corresponding location in a single write with no atomics. The kernel uses a grid-stride loop and read-only caching for high memory throughput and excellent performance on large feature maps.
\end{itemize}

\newpage
\section{Highlighted CUDA Kernels: Code}
\label{appsec:kernels}

\subsection{MNIST Conv-ReLU-Pool Forward Kernel}

\begin{lstlisting}[language=C++, caption=Conv-ReLU-Pool Fused CUDA Kernel, label=lst:conv_relu_pool_kernel, breaklines=true]
#include <torch/extension.h>
#include <cuda.h>
#include <cuda_runtime.h>

template<int K, int TX, int TY>
__global__ void unrolled_threadmapped_kernel(
    const float* __restrict__ input,   // [N, C_in, H_in, W_in]
    const float* __restrict__ weight,  // [C_out, C_in, K, K]
    const float* __restrict__ bias,    // [C_out]
    float* __restrict__ output,        // [N, C_out, H_out, W_out]
    int N, int C_in, int C_out,
    int H_in, int W_in) {

    int H_conv = H_in - K + 1, W_conv = W_in - K + 1;
    int H_out = H_conv >> 1, W_out = W_conv >> 1;

    int z = blockIdx.z;
    int n  = z / C_out;
    int oc = z % C_out;
    if (n >= N) return;

    int out_x0 = blockIdx.x * TX;
    int out_y0 = blockIdx.y * TY;

    extern __shared__ float shm_w[];
    int WperThread = (C_in*K*K + TX*TY - 1) / (TX*TY);
    int tid = threadIdx.y*TX + threadIdx.x;
    for(int i=0; i<WperThread; ++i){
        int idx = tid + i*TX*TY;
        if(idx < C_in*K*K){
            shm_w[idx] = weight[oc*C_in*K*K + idx];
        }
    }
    __syncthreads();

    int tx = threadIdx.x, ty = threadIdx.y;
    int ow = out_x0 + tx;
    int oh = out_y0 + ty;
    if(ow < W_out && oh < H_out){
        float best = -1e20f;
        #pragma unroll 2
        for(int ph=0; ph<2; ++ph){
            #pragma unroll 2
            for(int pw=0; pw<2; ++pw){
                int cv_y = (oh<<1)+ph;
                int cv_x = (ow<<1)+pw;
                float acc = bias[oc];
                for(int ic=0; ic<C_in; ++ic){
                    int base = ic*K*K;
                    #pragma unroll
                    for(int ky=0; ky<K; ++ky){
                        int in_y = cv_y+ky;
                        const float* row = input + ((n*C_in+ic)*H_in + in_y)*W_in;
                        #pragma unroll
                        for(int kx=0; kx<K; ++kx){
                            acc += row[cv_x + kx] * shm_w[base + ky*K + kx];
                        }
                    }
                }
                acc = acc>0.f? acc:0.f;
                best = acc>best? acc:best;
            }
        }
        int out_idx = ((n*C_out+oc)*H_out + oh)*W_out + ow;
        output[out_idx] = best;
    }
}

torch::Tensor forward_cuda(
    torch::Tensor input,
    torch::Tensor weight,
    torch::Tensor bias) {

    int N       = input.size(0);
    int C_in    = input.size(1);
    int H_in    = input.size(2);
    int W_in    = input.size(3);
    int C_out   = weight.size(0);
    int K       = weight.size(2);

    int H_conv  = H_in - K + 1;
    int W_conv  = W_in - K + 1;
    int H_out   = H_conv >> 1;
    int W_out   = W_conv >> 1;

    auto output = torch::empty({N, C_out, H_out, W_out}, input.options());

    constexpr int TX = 16, TY = 8;
    dim3 threads(TX, TY);
    int gx = (W_out + TX - 1)/TX;
    int gy = (H_out + TY - 1)/TY;
    dim3 blocks(gx, gy, N*C_out);

    size_t shm = C_in*K*K * sizeof(float);

    // Unroll for K=3 (as in MNIST)
    if (K == 3) {
        unrolled_threadmapped_kernel<3,TX,TY><<<blocks,threads,shm>>>(
            input.data_ptr<float>(),
            weight.data_ptr<float>(),
            bias.data_ptr<float>(),
            output.data_ptr<float>(),
            N,C_in,C_out,H_in,W_in
        );
    } else {
        // fallback: no unrolling
        unrolled_threadmapped_kernel<5,TX,TY><<<blocks,threads,shm>>>(
            input.data_ptr<float>(),
            weight.data_ptr<float>(),
            bias.data_ptr<float>(),
            output.data_ptr<float>(),
            N,C_in,C_out,H_in,W_in
        );
    }

    return output;
}

PYBIND11_MODULE(TORCH_EXTENSION_NAME, m) {
    m.def("forward", &forward_cuda, "Unrolled thread-mapped fused Conv2d+ReLU+MaxPool2d (CUDA)");
}
\end{lstlisting}

\subsection{MNIST Linear-ReLU Forward Kernel}

\begin{lstlisting}[language=C++, caption=Linear-ReLU Fused CUDA Kernel, label=lst:linear_relu, breaklines=true]
#include <torch/extension.h>
#include <cuda.h>
#include <cuda_runtime.h>
#include <vector>

#define WARP_SIZE 32
#define WARPS_PER_BLOCK 4  // blockDim.x = 128

template<typename scalar_t>
__global__ void linear_relu_shfl_strideloop_kernel(
    const scalar_t* __restrict__ input,    // [B, I]
    const scalar_t* __restrict__ weights,  // [O, I]
    const scalar_t* __restrict__ bias,     // [O]
    scalar_t* __restrict__ output,         // [B, O]
    int B, int I, int O)
{
    int threads_per_block = blockDim.x;
    int blocks_x = gridDim.x;
    int blocks_y = gridDim.y;

    int warps_per_block = threads_per_block / WARP_SIZE;
    int global_warp_id = (blockIdx.y * blocks_x + blockIdx.x) * warps_per_block + threadIdx.x / WARP_SIZE;
    int total_warps = blocks_x * blocks_y * warps_per_block;

    int lane_id = threadIdx.x & (WARP_SIZE - 1);

    // Stride loop over (batch, out_feature) pairs, each warp computes one output per loop
    for (int idx = global_warp_id; idx < B * O; idx += total_warps) {
        int batch_idx = idx / O;
        int out_idx   = idx % O;

        if (batch_idx < B && out_idx < O) {
            const scalar_t* x_row = input  + batch_idx * I;
            const scalar_t* w_row = weights + out_idx   * I;

            // Each lane computes partial dot-product
            scalar_t sum = scalar_t(0);
            for (int k = lane_id; k < I; k += WARP_SIZE) {
                sum += x_row[k] * w_row[k];
            }

            // Warp-wide reduction using shuffle
            unsigned full_mask = 0xffffffffu;
            for (int offset = WARP_SIZE / 2; offset > 0; offset >>= 1) {
                sum += __shfl_down_sync(full_mask, sum, offset);
            }

            // Warp leader writes result
            if (lane_id == 0) {
                scalar_t val = sum + bias[out_idx];
                output[batch_idx * O + out_idx] = val > scalar_t(0) ? val : scalar_t(0);
            }
        }
    }
}

torch::Tensor linear_relu_shfl_strideloop_forward(
    torch::Tensor input,    // [B, I]
    torch::Tensor weights,  // [O, I]
    torch::Tensor bias)     // [O]
{
    TORCH_CHECK(input.is_contiguous(),  "input must be contiguous");
    TORCH_CHECK(weights.is_contiguous(),"weights must be contiguous");
    TORCH_CHECK(bias.is_contiguous(),   "bias must be contiguous");
    int B = input.size(0);
    int I = input.size(1);
    int O = weights.size(0);

    auto output = torch::empty({B, O}, input.options());

    // Launch configuration: each block has WARPS_PER_BLOCK warps
    int threads = WARP_SIZE * WARPS_PER_BLOCK;
    // Use a 2D grid for flexibility and occupancy
    int max_blocks = 512; // tune for your device
    int blocks_x = std::min((O + WARPS_PER_BLOCK - 1) / WARPS_PER_BLOCK, max_blocks);
    int blocks_y = std::min((B + 1), max_blocks);

    dim3 grid(blocks_x, blocks_y);
    dim3 block(threads);

    AT_DISPATCH_FLOATING_TYPES(input.scalar_type(), "linear_relu_shfl_strideloop_forward", ([&] {
        linear_relu_shfl_strideloop_kernel<scalar_t><<<grid, block>>>(
            input.data_ptr<scalar_t>(),
            weights.data_ptr<scalar_t>(),
            bias.data_ptr<scalar_t>(),
            output.data_ptr<scalar_t>(),
            B, I, O);
    }));

    cudaError_t err = cudaGetLastError();
    if (err != cudaSuccess) {
        throw std::runtime_error("CUDA error: " + std::string(cudaGetErrorString(err)));
    }
    return output;
}

PYBIND11_MODULE(TORCH_EXTENSION_NAME, m) {
    m.def("forward", &linear_relu_shfl_strideloop_forward, "Linear+ReLU with warp-shuffle and grid-stride loop");
}
\end{lstlisting}

\subsection{MNIST Linear Forward Kernel}

\begin{lstlisting}[language=C++, caption=Linear CUDA Kernel, label=lst:linear, breaklines=true]
#include <torch/extension.h>
#include <cuda_runtime.h>

#define WARP_SIZE 32

// Each block (one warp) computes y[row, col].
// Threads in the warp cooperatively load and multiply0add elements,
// then reduce via warp-shuffle, and thread 0 adds bias and writes output.
__global__ void linear_forward_warp_kernel(
    const float* __restrict__ x,
    const float* __restrict__ weights,
    const float* __restrict__ bias,
    float* __restrict__ y,
    int m,
    int n,
    int k
) {
    int col = blockIdx.x;
    int row = blockIdx.y;
    if (row >= m || col >= n) return;

    int lane = threadIdx.x;  // [0..31]
    float acc = 0.0f;

    // each lane accumulates x[row,k] * weights[col,k] over k in strides of warpSize
    for (int p = lane; p < k; p += WARP_SIZE) {
        acc += x[row * k + p] * weights[col * k + p];
    }

    // warp-level reduction using shuffle down
    #pragma unroll
    for (int offset = WARP_SIZE/2; offset > 0; offset >>= 1) {
        acc += __shfl_down_sync(0xffffffff, acc, offset);
    }

    // lane 0 writes the final result + bias
    if (lane == 0) {
        y[row * n + col] = acc + bias[col];
    }
}

torch::Tensor forward(torch::Tensor x, torch::Tensor weights, torch::Tensor bias) {
    x = x.contiguous();
    weights = weights.contiguous();
    bias = bias.contiguous();

    int m = x.size(0);
    int k = x.size(1);
    int n = weights.size(0);

    auto y = torch::empty({m, n}, x.options());

    dim3 block(WARP_SIZE);
    dim3 grid(n, m);
    linear_forward_warp_kernel<<<grid, block>>>(
        x.data_ptr<float>(),
        weights.data_ptr<float>(),
        bias.data_ptr<float>(),
        y.data_ptr<float>(),
        m, n, k
    );
    return y;
}

PYBIND11_MODULE(TORCH_EXTENSION_NAME, m) {
    m.def("forward", &forward, "Linear layer forward pass using warp-shuffle (CUDA)");
}
\end{lstlisting}

\subsection{MNIST Cross-Entropy Forward Kernel}

\begin{lstlisting}[language=C++, caption=Cross-Entropy CUDA Kernel, label=lst:ce, breaklines=true]
#include <torch/extension.h>
#include <ATen/cuda/CUDAContext.h>
#include <cuda.h>
#include <cuda_runtime.h>

#define THREADS_PER_BLOCK 256
#define WARP_SIZE 32
#define SAMPLES_PER_BLOCK (THREADS_PER_BLOCK / WARP_SIZE)

// Template kernel for known class count (unrolled)
template<int NUM_CLASSES>
__global__ void cross_entropy_sharedwarp_unroll_kernel(
    const float* __restrict__ preds,
    const int64_t* __restrict__ targets,
    float* __restrict__ loss,
    int num_samples)
{
    extern __shared__ float shmem[];
    int tid      = threadIdx.x;
    int warp_id  = tid / WARP_SIZE;
    int lane     = tid % WARP_SIZE;
    int sample_i = blockIdx.x * SAMPLES_PER_BLOCK + warp_id;

    if (sample_i < num_samples) {
        const float* row_g = preds + sample_i * NUM_CLASSES;
        float* row_s = shmem + warp_id * NUM_CLASSES;

        // 1) Load row into shared memory (unrolled)
        #pragma unroll
        for (int j = lane; j < NUM_CLASSES; j += WARP_SIZE) {
            row_s[j] = row_g[j];
        }
        __syncwarp();

        // 2) Find max (unrolled)
        float local_max = -1e30f;
        #pragma unroll
        for (int j = lane; j < NUM_CLASSES; j += WARP_SIZE) {
            local_max = fmaxf(local_max, row_s[j]);
        }
        #pragma unroll
        for (int offset = WARP_SIZE/2; offset > 0; offset >>= 1) {
            float m = __shfl_down_sync(0xffffffff, local_max, offset);
            local_max = fmaxf(local_max, m);
        }
        float max_val = __shfl_sync(0xffffffff, local_max, 0);

        // 3) Sum exp (unrolled)
        float local_sum = 0.0f;
        #pragma unroll
        for (int j = lane; j < NUM_CLASSES; j += WARP_SIZE) {
            local_sum += __expf(row_s[j] - max_val);
        }
        #pragma unroll
        for (int offset = WARP_SIZE/2; offset > 0; offset >>= 1) {
            float s = __shfl_down_sync(0xffffffff, local_sum, offset);
            local_sum += s;
        }
        float sum_exp = __shfl_sync(0xffffffff, local_sum, 0);

        int64_t t = targets[sample_i];
        float tval = row_s[t];

        float logp = tval - max_val - __logf(sum_exp);
        if (lane == 0) {
            atomicAdd(loss, -logp);
        }
    }
}

// Fallback kernel for unknown/large class count (partial unrolling)
__global__ void cross_entropy_sharedwarp_unroll_kernel_dynamic(
    const float* __restrict__ preds,
    const int64_t* __restrict__ targets,
    float* __restrict__ loss,
    int num_samples,
    int num_classes)
{
    extern __shared__ float shmem[];
    int tid      = threadIdx.x;
    int warp_id  = tid / WARP_SIZE;
    int lane     = tid % WARP_SIZE;
    int sample_i = blockIdx.x * SAMPLES_PER_BLOCK + warp_id;

    if (sample_i < num_samples) {
        const float* row_g = preds + sample_i * num_classes;
        float* row_s = shmem + warp_id * num_classes;

        // 1) Load row into shared memory
        #pragma unroll 8
        for (int j = lane; j < num_classes; j += WARP_SIZE) {
            row_s[j] = row_g[j];
        }
        __syncwarp();

        // 2) Find max
        float local_max = -1e30f;
        #pragma unroll 8
        for (int j = lane; j < num_classes; j += WARP_SIZE) {
            local_max = fmaxf(local_max, row_s[j]);
        }
        #pragma unroll
        for (int offset = WARP_SIZE/2; offset > 0; offset >>= 1) {
            float m = __shfl_down_sync(0xffffffff, local_max, offset);
            local_max = fmaxf(local_max, m);
        }
        float max_val = __shfl_sync(0xffffffff, local_max, 0);

        // 3) Sum exp
        float local_sum = 0.0f;
        #pragma unroll 8
        for (int j = lane; j < num_classes; j += WARP_SIZE) {
            local_sum += __expf(row_s[j] - max_val);
        }
        #pragma unroll
        for (int offset = WARP_SIZE/2; offset > 0; offset >>= 1) {
            float s = __shfl_down_sync(0xffffffff, local_sum, offset);
            local_sum += s;
        }
        float sum_exp = __shfl_sync(0xffffffff, local_sum, 0);

        int64_t t = targets[sample_i];
        float tval = row_s[t];

        float logp = tval - max_val - __logf(sum_exp);
        if (lane == 0) {
            atomicAdd(loss, -logp);
        }
    }
}

// Host wrapper dispatches the correct kernel for common class counts
torch::Tensor forward(torch::Tensor predictions, torch::Tensor targets) {
    TORCH_CHECK(predictions.is_cuda(), "predictions must be CUDA");
    TORCH_CHECK(targets.is_cuda(),     "targets must be CUDA");
    TORCH_CHECK(predictions.dim()==2,   "predictions must be 2D");
    TORCH_CHECK(targets.dim()==1,       "targets must be 1D");
    TORCH_CHECK(predictions.size(0)==targets.size(0),
                "batch size mismatch");

    int N = predictions.size(0);
    int C = predictions.size(1);
    int threads = THREADS_PER_BLOCK;
    int blocks  = (N + SAMPLES_PER_BLOCK - 1) / SAMPLES_PER_BLOCK;

    auto loss_tensor = torch::zeros({1}, predictions.options());
    float* d_loss = loss_tensor.data_ptr<float>();

    const float*  d_preds  = predictions.data_ptr<float>();
    const int64_t* d_tgts  = targets.data_ptr<int64_t>();

    size_t shmem_bytes = size_t(SAMPLES_PER_BLOCK) * C * sizeof(float);

    // Specialize for common MNIST case (C==10), otherwise fallback
    if (C == 10) {
        cross_entropy_sharedwarp_unroll_kernel<10>
            <<<blocks, threads, shmem_bytes, at::cuda::getCurrentCUDAStream()>>>(
                d_preds, d_tgts, d_loss, N);
    } else if (C == 100) {
        cross_entropy_sharedwarp_unroll_kernel<100>
            <<<blocks, threads, shmem_bytes, at::cuda::getCurrentCUDAStream()>>>(
                d_preds, d_tgts, d_loss, N);
    } else {
        cross_entropy_sharedwarp_unroll_kernel_dynamic
            <<<blocks, threads, shmem_bytes, at::cuda::getCurrentCUDAStream()>>>(
                d_preds, d_tgts, d_loss, N, C);
    }
    return loss_tensor / static_cast<float>(N);
}

PYBIND11_MODULE(TORCH_EXTENSION_NAME, m) {
    m.def("forward", &forward, "CrossEntropy (shared-mem warp, unrolled) forward (CUDA)");
}
\end{lstlisting}

\subsection{MNIST MaxPool Backward Kernel}

\begin{lstlisting}[language=C++, caption=MaxPool Backward CUDA Kernel, label=lst:maxpool_backward, breaklines=true]
#include <torch/extension.h>
#include <pybind11/pybind11.h>
#include <ATen/ATen.h>
#include <ATen/cuda/CUDAContext.h>
#include <cuda.h>
#include <cuda_runtime.h>

namespace py = pybind11;

/* -------------------------- device kernel -------------------------------- */

template <typename scalar_t>
__global__ void maxpool2d_ks2_backward_fast_kernel(
        const scalar_t* __restrict__ x,
        const scalar_t* __restrict__ grad_out,
              scalar_t* __restrict__ grad_in,
        int N, int C, int H, int W,
        int outH, int outW,
        long total_out)               // N*C*outH*outW
{
    // grid-stride loop - lets us choose any grid size
    for (long idx = blockIdx.x * blockDim.x + threadIdx.x;
         idx < total_out;
         idx += gridDim.x * blockDim.x)
    {
        // unravel linear index ->  n, c, h_out, w_out
        int w_out = idx % outW;
        int tmp   = idx / outW;
        int h_out = tmp % outH;
        tmp       = tmp / outH;
        int c     = tmp % C;
        int n     = tmp / C;

        // pointer to the plane (n,c)
        const long plane_offset_X  = ((long)n * C + c) * H * W;
        const long plane_offset_GO = ((long)n * C + c) * outH * outW;

        // top-left of the 2x2 window in input
        int h_in = h_out << 1;    // *2
        int w_in = w_out << 1;

        const scalar_t* x_ptr = x + plane_offset_X + (long)h_in * W + w_in;

        // load the four values (use read-only cache)
        scalar_t v00 = __ldg(x_ptr);           // (0,0)
        scalar_t v01 = __ldg(x_ptr + 1);       // (0,1)
        scalar_t v10 = __ldg(x_ptr + W);       // (1,0)
        scalar_t v11 = __ldg(x_ptr + W + 1);   // (1,1)

        // locate the maximum
        scalar_t maxv = v00;  int offset = 0;
        if (v01 > maxv) { maxv = v01; offset = 1; }
        if (v10 > maxv) { maxv = v10; offset =  W; }
        if (v11 > maxv) { maxv = v11; offset =  W + 1; }

        // route upstream gradient (grad_input is zero-initialised -> 1 write)
        grad_in[plane_offset_X + (long)h_in * W + w_in + offset] =
            grad_out[plane_offset_GO + (long)h_out * outW + w_out];
    }
}

/* --------------------------- launcher ------------------------------------ */

at::Tensor maxpool2d_ks2_backward_fast(
        const at::Tensor& x,
        const at::Tensor& grad_out)
{
    TORCH_CHECK(x.is_cuda() && grad_out.is_cuda(),
                "tensors must be CUDA");

    const int N = x.size(0);
    const int C = x.size(1);
    const int H = x.size(2);
    const int W = x.size(3);
    const int outH = grad_out.size(2);
    const int outW = grad_out.size(3);

    // grad_input initialised to zero once; no extra writes in the kernel
    auto grad_in = at::zeros_like(x);

    const long total = 1L * N * C * outH * outW;

    /* kernel configuration:
       - 256 threads / block gives good occupancy on H100
       - blocks = min( ceil(total/threads),  65535 )             */
    const int threads = 256;
    const int maxBlocks = 65535;
    int blocks = (int)std::min( (total + threads - 1) / threads,
                                (long)maxBlocks );

    cudaStream_t stream = at::cuda::getCurrentCUDAStream();

    AT_DISPATCH_FLOATING_TYPES_AND_HALF(
        x.scalar_type(), "maxpool2d_ks2_backward_fast", [&]
    {
        maxpool2d_ks2_backward_fast_kernel<scalar_t>
            <<<blocks, threads, 0, stream>>>(
                x.data_ptr<scalar_t>(),
                grad_out.data_ptr<scalar_t>(),
                grad_in.data_ptr<scalar_t>(),
                N, C, H, W, outH, outW, total);
    });

    return grad_in;
}

/* --------------------------- python glue --------------------------------- */

at::Tensor backward_wrapper(py::tuple saved, const at::Tensor& grad_out)
{
    TORCH_CHECK(saved.size() == 1,
                "expected a single saved tensor");
    at::Tensor x = saved[0].cast<at::Tensor>();
    return maxpool2d_ks2_backward_fast(x, grad_out);
}

PYBIND11_MODULE(TORCH_EXTENSION_NAME, m)
{
    m.def("backward", &backward_wrapper,
          "2x2-stride-2 max-pool backward (fast, no atomics)");
}
\end{lstlisting}

\subsection{MNIST Linear-ReLU Backward Kernel}

\begin{lstlisting}[language=C++, caption=Linear-ReLU Backward CUDA Kernel, label=lst:linear_relu_backward, breaklines=true]
#include <torch/extension.h>
#include <cuda.h>
#include <cuda_runtime.h>
#include <vector>

#define CHECK_TENSOR(x) TORCH_CHECK(x.is_cuda() && x.is_contiguous(), #x " must be a contiguous CUDA tensor")

// -----------------------------------------------------------------------------
// kernel 1 : build masked gradient and bias gradient
// -----------------------------------------------------------------------------
__global__ void masked_grad_bias_kernel(const float *__restrict__ x,
                                        const float *__restrict__ w,
                                        const float *__restrict__ bias,
                                        const float *__restrict__ grad_out,
                                        float *__restrict__ masked_grad,
                                        float *__restrict__ grad_bias,
                                        const int batch,
                                        const int in_feat,
                                        const int out_feat)
{
    const int o = blockIdx.x;                                // output neuron handled by this block
    const int tid = threadIdx.x;
    const int stride = blockDim.x;

    for (int b = tid; b < batch; b += stride)
    {
        // dot( x[b], w[o] )
        float acc = 0.f;
        const float *x_ptr = x + b * in_feat;
        const float *w_ptr = w + o * in_feat;
#pragma unroll
        for (int i = 0; i < in_feat; ++i)
            acc += x_ptr[i] * w_ptr[i];

        acc += bias[o];                                      // + b

        const float go = grad_out[b * out_feat + o];
        const float mg = (acc > 0.f) ? go : 0.f;             // apply mask

        masked_grad[b * out_feat + o] = mg;                  // store masked gradient

        atomicAdd(grad_bias + o, mg);                        // accumulate bias gradient
    }
}

// -----------------------------------------------------------------------------
// kernel 2 : weight gradient  (dL/dW = masked_grad^T @ x)
// one block  -> one output neuron
// one thread -> several input features (looped by stride if needed)
// -----------------------------------------------------------------------------
__global__ void grad_weights_kernel(const float *__restrict__ x,
                                    const float *__restrict__ masked_grad,
                                    float *__restrict__ grad_w,
                                    const int batch,
                                    const int in_feat,
                                    const int out_feat)
{
    const int o = blockIdx.x;               // output neuron for this block
    const int tid = threadIdx.x;
    const int stride = blockDim.x;

    for (int i = tid; i < in_feat; i += stride)
    {
        float acc = 0.f;
        for (int b = 0; b < batch; ++b)
            acc += x[b * in_feat + i] * masked_grad[b * out_feat + o];

        grad_w[o * in_feat + i] = acc;      // write gradient
    }
}

// -----------------------------------------------------------------------------
// kernel 3 : input gradient (dL/dx = masked_grad @ W)
// one block  -> one sample
// one thread -> several input features (looped by stride if needed)
// -----------------------------------------------------------------------------
__global__ void grad_input_kernel(const float *__restrict__ masked_grad,
                                  const float *__restrict__ w,
                                  float *__restrict__ grad_in,
                                  const int batch,
                                  const int in_feat,
                                  const int out_feat)
{
    const int b = blockIdx.x;               // sample handled by this block
    const int tid = threadIdx.x;
    const int stride = blockDim.x;

    for (int i = tid; i < in_feat; i += stride)
    {
        float acc = 0.f;
        const float *w_col = w + i;         // column i across all out neurons
#pragma unroll
        for (int o = 0; o < out_feat; ++o)
            acc += masked_grad[b * out_feat + o] *
                   w_col[o * in_feat];

        grad_in[b * in_feat + i] = acc;
    }
}

// -----------------------------------------------------------------------------
// Host dispatcher
// -----------------------------------------------------------------------------
std::vector<torch::Tensor> backward(torch::Tensor grad_out,
                                    torch::Tensor x,
                                    torch::Tensor weight,
                                    torch::Tensor bias)
{
    CHECK_TENSOR(grad_out);
    CHECK_TENSOR(x);
    CHECK_TENSOR(weight);
    CHECK_TENSOR(bias);

    const int batch      = x.size(0);
    const int in_feat    = x.size(1);
    const int out_feat   = weight.size(0);

    auto grad_input   = torch::empty_like(x);
    auto grad_weight  = torch::empty_like(weight);
    auto grad_bias    = torch::zeros_like(bias);          // zero-ed for atomic adds
    auto masked_grad  = torch::empty_like(grad_out);      // temporary buffer

    // launch parameters
    constexpr int THREADS = 256;

    // 1) masked grad + bias grad
    masked_grad_bias_kernel<<<out_feat, THREADS>>>(x.data_ptr<float>(),
                                                   weight.data_ptr<float>(),
                                                   bias.data_ptr<float>(),
                                                   grad_out.data_ptr<float>(),
                                                   masked_grad.data_ptr<float>(),
                                                   grad_bias.data_ptr<float>(),
                                                   batch,
                                                   in_feat,
                                                   out_feat);

    // 2) weight grad
    grad_weights_kernel<<<out_feat, THREADS>>>(x.data_ptr<float>(),
                                               masked_grad.data_ptr<float>(),
                                               grad_weight.data_ptr<float>(),
                                               batch,
                                               in_feat,
                                               out_feat);

    // 3) input grad
    grad_input_kernel<<<batch, THREADS>>>(masked_grad.data_ptr<float>(),
                                          weight.data_ptr<float>(),
                                          grad_input.data_ptr<float>(),
                                          batch,
                                          in_feat,
                                          out_feat);

    return {grad_input, grad_weight, grad_bias};
}

PYBIND11_MODULE(TORCH_EXTENSION_NAME, m)
{
    m.def("backward", &backward, "Fused Linear+ReLU backward (x @ W^T + b)");
}
\end{lstlisting}

\subsection{MNIST Linear Backward Kernel}

\begin{lstlisting}[language=C++, caption=Linear Backward CUDA Kernel, label=lst:linear_backward, breaklines=true]
#include <torch/extension.h>
#include <ATen/cuda/CUDAContext.h>
#include <cuda.h>
#include <cuda_runtime.h>
#include <c10/cuda/CUDAMathCompat.h>

// Block size settings (can be tuned)
#define BLOCK_GRAD_INPUT 128
#define BLOCKX_WEIGHT 64
#define BLOCKY_WEIGHT 4
#define BLOCK_BIAS 256

// 1D row-parallel kernel for grad_input: out = grad_output @ weights^T
template <typename scalar_t>
__global__ void grad_input_adaptive_kernel(
    const scalar_t* __restrict__ grad_output,  // (B, O)
    const scalar_t* __restrict__ weights,      // (O, I)
    scalar_t* __restrict__ grad_input,         // (B, I)
    int B, int I, int O) {

    int row = blockIdx.y;
    int tx = threadIdx.x;
    for (int col = blockIdx.x * BLOCK_GRAD_INPUT + tx; col < I; col += blockDim.x * gridDim.x) {
        if (row >= B || col >= I) return;
        scalar_t sum = 0;
        // Unroll by 4
        int k = 0;
        #pragma unroll
        for (; k + 3 < O; k += 4)
            sum += grad_output[row * O + k    ] * weights[k    * I + col] +
                   grad_output[row * O + k + 1] * weights[(k + 1) * I + col] +
                   grad_output[row * O + k + 2] * weights[(k + 2) * I + col] +
                   grad_output[row * O + k + 3] * weights[(k + 3) * I + col];
        for (; k < O; ++k)
            sum += grad_output[row * O + k] * weights[k * I + col];
        grad_input[row * I + col] = sum;
    }
}

// 2D block kernel for grad_weights: out = grad_output^T @ x
template <typename scalar_t>
__global__ void grad_weights_adaptive_kernel(
    const scalar_t* __restrict__ grad_output,  // (B, O)
    const scalar_t* __restrict__ x,            // (B, I)
    scalar_t* __restrict__ grad_weights,       // (O, I)
    int B, int O, int I) {

    int o = blockIdx.y * BLOCKY_WEIGHT + threadIdx.y;
    int i = blockIdx.x * BLOCKX_WEIGHT + threadIdx.x;
    if (o >= O || i >= I) return;
    scalar_t sum = 0;
    int b = 0;
    #pragma unroll
    for (; b + 3 < B; b += 4)
        sum += grad_output[(b    ) * O + o] * x[(b    ) * I + i] +
               grad_output[(b + 1) * O + o] * x[(b + 1) * I + i] +
               grad_output[(b + 2) * O + o] * x[(b + 2) * I + i] +
               grad_output[(b + 3) * O + o] * x[(b + 3) * I + i];
    for (; b < B; ++b)
        sum += grad_output[b * O + o] * x[b * I + i];
    grad_weights[o * I + i] = sum;
}

// bias reduction kernel: block per output feature, 256 threads per block
template <typename scalar_t>
__global__ void grad_bias_adaptive_kernel(
    const scalar_t* __restrict__ grad_output,  // (B, O)
    scalar_t* __restrict__ grad_bias,          // (O)
    int B, int O) {

    int o = blockIdx.x;
    __shared__ scalar_t sdata[BLOCK_BIAS];
    scalar_t sum = 0;
    for (int b = threadIdx.x; b < B; b += blockDim.x)
        sum += grad_output[b * O + o];
    sdata[threadIdx.x] = sum;
    __syncthreads();

    // block-wide reduction
    for (int stride = BLOCK_BIAS / 2; stride > 0; stride /= 2) {
        if (threadIdx.x < stride)
            sdata[threadIdx.x] += sdata[threadIdx.x + stride];
        __syncthreads();
    }
    if (threadIdx.x == 0)
        grad_bias[o] = sdata[0];
}

std::vector<at::Tensor> backward(
    at::Tensor grad_output,
    at::Tensor x,
    at::Tensor weights) {
    int B = grad_output.size(0);
    int O = grad_output.size(1);
    int I = x.size(1);

    auto grad_input = at::empty({B, I}, grad_output.options());
    auto grad_weights = at::empty({O, I}, grad_output.options());
    auto grad_bias = at::empty({O}, grad_output.options());

    // grad_input: block per row, each block processes BLOCK_GRAD_INPUT columns
    dim3 block_input(BLOCK_GRAD_INPUT);
    dim3 grid_input((I + BLOCK_GRAD_INPUT - 1) / BLOCK_GRAD_INPUT, B);

    AT_DISPATCH_FLOATING_TYPES(grad_output.scalar_type(), "grad_input_adaptive_kernel", ([&] {
        grad_input_adaptive_kernel<scalar_t><<<grid_input, block_input, 0, at::cuda::getCurrentCUDAStream()>>>(
            grad_output.data_ptr<scalar_t>(),
            weights.data_ptr<scalar_t>(),
            grad_input.data_ptr<scalar_t>(),
            B, I, O);
    }));

    // grad_weights: 2D grid
    dim3 block_weight(BLOCKX_WEIGHT, BLOCKY_WEIGHT);
    dim3 grid_weight((I + BLOCKX_WEIGHT - 1) / BLOCKX_WEIGHT,
                    (O + BLOCKY_WEIGHT - 1) / BLOCKY_WEIGHT);

    AT_DISPATCH_FLOATING_TYPES(grad_output.scalar_type(), "grad_weights_adaptive_kernel", ([&] {
        grad_weights_adaptive_kernel<scalar_t><<<grid_weight, block_weight, 0, at::cuda::getCurrentCUDAStream()>>>(
            grad_output.data_ptr<scalar_t>(),
            x.data_ptr<scalar_t>(),
            grad_weights.data_ptr<scalar_t>(),
            B, O, I);
    }));

    // grad_bias: block per output feature, 256 threads per block
    dim3 block_bias(BLOCK_BIAS);
    dim3 grid_bias(O);

    AT_DISPATCH_FLOATING_TYPES(grad_output.scalar_type(), "grad_bias_adaptive_kernel", ([&] {
        grad_bias_adaptive_kernel<scalar_t><<<grid_bias, block_bias, 0, at::cuda::getCurrentCUDAStream()>>>(
            grad_output.data_ptr<scalar_t>(),
            grad_bias.data_ptr<scalar_t>(),
            B, O);
    }));

    return {grad_input, grad_weights, grad_bias};
}

PYBIND11_MODULE(TORCH_EXTENSION_NAME, m) {
    m.def("backward", &backward, "Adaptive block size linear backward CUDA kernel");
}
\end{lstlisting}

\subsection{MNIST Cross-Entropy Backward Kernel}

\lstinputlisting[language=C++, caption=Cross-Entropy Backward CUDA Kernel, label=lst:ce_backward, breaklines=true]{supps/highlighted/mnist_cross_entropy/backward/kernel.cu}

\subsection{LayerNorm Forward Kernel}
\label{appsec:layernorm_kernel}

\lstinputlisting[language=C++, caption=LayerNorm Fused Vectorized CUDA Kernel, label=lst:layernorm_kernel, breaklines=true]{supps/highlighted/layernorm/forward/kernel.cu}

\subsection{ResNet Block Forward Kernel}

\lstinputlisting[language=C++, caption=ResNet Block Fused CUDA Kernel, label=lst:resnet_block_kernel, breaklines=true]{supps/highlighted/resnet_block/forward/kernel.cu}

\subsection{Llama RMSNorm Forward Kernel}

\begin{lstlisting}[language=C++, caption=Llama RMSNorm Fused CUDA Kernel, label=lst:llama_rmsnorm_kernel, breaklines=true]
#include <torch/extension.h>
#include <ATen/cuda/CUDAContext.h>
#include <cuda.h>
#include <cuda_runtime.h>

#define WARP_SIZE 32

template<typename T>
__device__ __forceinline__ T warp_sum(T v) {
    #pragma unroll
    for (int d = WARP_SIZE / 2; d > 0; d >>= 1)
        v += __shfl_down_sync(0xffffffff, v, d);
    return v;
}

template<int BLOCK_SIZE>
__global__ void strided_rms_norm_kernel(
    const float* __restrict__ x,
    const float* __restrict__ w,
    float eps,
    float* __restrict__ y,
    int M,
    int D
) {
    extern __shared__ float s_partials[];
    int row = blockIdx.x;
    if (row >= M) return;
    const float* row_x = x + row * D;
    float* row_y = y + row * D;

    // 1. Compute sum of squares using a strided loop
    float sum = 0.f;
    int tid = threadIdx.x;
    int nthreads = blockDim.x;

    // Vectorized float4 loads where possible
    int D4 = D / 4 * 4;
    for (int i = tid * 4; i < D4; i += nthreads * 4) {
        float4 v4 = reinterpret_cast<const float4*>(row_x)[i / 4];
        sum += v4.x * v4.x + v4.y * v4.y + v4.z * v4.z + v4.w * v4.w;
    }
    for (int i = D4 + tid; i < D; i += nthreads) {
        float v = row_x[i];
        sum += v * v;
    }

    // Block-wide reduction
    sum = warp_sum(sum);
    if ((tid & (WARP_SIZE - 1)) == 0)
        s_partials[tid / WARP_SIZE] = sum;
    __syncthreads();

    float total = 0.f;
    if (tid < nthreads / WARP_SIZE)
        total = s_partials[tid];
    if (tid < WARP_SIZE)
        total = warp_sum(total);
    float norm_factor = 0.f;
    if (tid == 0)
        s_partials[0] = rsqrtf(total / D + eps);
    __syncthreads();
    norm_factor = s_partials[0];

    // 2. Normalize and write out using strided loop (float4 where possible)
    for (int i = tid * 4; i < D4; i += nthreads * 4) {
        float4 v4 = reinterpret_cast<const float4*>(row_x)[i / 4];
        float4 w4 = reinterpret_cast<const float4*>(w)[i / 4];
        float4 y4;
        y4.x = v4.x * norm_factor * w4.x;
        y4.y = v4.y * norm_factor * w4.y;
        y4.z = v4.z * norm_factor * w4.z;
        y4.w = v4.w * norm_factor * w4.w;
        reinterpret_cast<float4*>(row_y)[i / 4] = y4;
    }
    for (int i = D4 + tid; i < D; i += nthreads) {
        row_y[i] = row_x[i] * norm_factor * w[i];
    }
}

static inline int pick_block_size(int D) {
    if (D >= 4096) return 512;
    if (D >= 2048) return 256;
    if (D >= 1024) return 128;
    if (D >= 512)  return 64;
    return 32;
}

template<int BS>
void launch_strided(const float* x, const float* w, float eps, float* y,
                    int M, int D, cudaStream_t s)
{
    size_t smem = (BS / WARP_SIZE) * sizeof(float);
    strided_rms_norm_kernel<BS><<<M, BS, smem, s>>>(x, w, eps, y, M, D);
}

at::Tensor forward(at::Tensor x, at::Tensor w, double eps) {
    TORCH_CHECK(x.is_cuda() && w.is_cuda(), "tensors must be CUDA");
    TORCH_CHECK(x.is_contiguous() && w.is_contiguous(), "tensors must be contiguous");
    TORCH_CHECK(x.scalar_type() == at::kFloat, "only float32 supported");

    int D = x.size(-1);
    int M = x.numel() / D;
    TORCH_CHECK(w.size(0) == D, "weight size mismatch");

    at::Tensor y = at::empty_like(x);
    const float* x_ptr = x.data_ptr<float>();
    const float* w_ptr = w.data_ptr<float>();
    float* y_ptr = y.data_ptr<float>();

    int block = pick_block_size(D);
    cudaStream_t stream = at::cuda::getCurrentCUDAStream();

    switch (block) {
        case 512: launch_strided<512>(x_ptr, w_ptr, static_cast<float>(eps), y_ptr, M, D, stream); break;
        case 256: launch_strided<256>(x_ptr, w_ptr, static_cast<float>(eps), y_ptr, M, D, stream); break;
        case 128: launch_strided<128>(x_ptr, w_ptr, static_cast<float>(eps), y_ptr, M, D, stream); break;
        case 64:  launch_strided<64 >(x_ptr, w_ptr, static_cast<float>(eps), y_ptr, M, D, stream); break;
        default:  launch_strided<32 >(x_ptr, w_ptr, static_cast<float>(eps), y_ptr, M, D, stream); break;
    }

    cudaError_t err = cudaGetLastError();
    TORCH_CHECK(err == cudaSuccess, cudaGetErrorString(err));
    return y;
}

PYBIND11_MODULE(TORCH_EXTENSION_NAME, m) {
    m.def("forward", &forward, "RMSNorm with strided loops for large workloads (CUDA)");
}
\end{lstlisting}

\subsection{Llama Feedforward Block Forward Kernel}

\lstinputlisting[language=C++, caption=Llama Feedforward Block Fused CUDA Kernel, label=lst:llama_feedforward_kernel, breaklines=true]{supps/highlighted/llama_ffw/forward/kernel.cu}